\documentclass[]{aastex7}

\usepackage{starfont}

\def\lya{\mbox{Ly$\alpha$}}

\usepackage{amsmath,amssymb}

\usepackage{caption}
\usepackage{subcaption}

\begin{document}

\title{Lyman-Alpha Emission from K and M Dwarfs: Intrinsic Profiles, Variability, and Flux in the Habitable Zone}
 
\author[0000-0002-1046-025X]{Sarah Peacock} 
\affiliation{University of Maryland, Baltimore County, MD 21250, USA} 
\affiliation{NASA Goddard Space Flight Center, Greenbelt, MD 20771, USA}
\email{sarah.r.peacock@nasa.gov}

\author[0000-0002-7129-3002]{Travis S. Barman}
\affiliation{University of Arizona, Lunar and Planetary Laboratory, 1629 E University Boulevard, Tucson, AZ 85721, USA}
\email{barman@lpl.arizona.edu}

\author[0000-0001-5646-6668]{R. O. Parke Loyd}
\affil{Eureka Scientific, Inc., 2452 Delmer Street Suite 100, Oakland, CA 94602-3017, USA}
\email{astroparke@gmail.com}

\author[0000-0002-6294-5937]{Adam C. Schneider}
\affil{US Naval Observatory, Flagstaff Station, 10391 West Naval Observatory Road, Flagstaff, AZ 86002-8521, USA}
\email{aschneid10@gmail.com}

\author[0000-0002-1176-3391]{Allison Youngblood}
\affiliation{NASA Goddard Space Flight Center, Greenbelt, MD 20771, USA}
\email{allison.a.youngblood@nasa.gov}

\author[0000-0002-0706-079X]{Kenneth G. Carpenter} 
\affiliation{NASA Goddard Space Flight Center, Greenbelt, MD 20771, USA}
\email{kenneth.g.carpenter@nasa.gov}

\author[0000-0002-7260-5821]{Evgenya L. Shkolnik}
\affil{School of Earth and Space Exploration, Arizona State University, Tempe, AZ 85281, USA}
\email{eshkolni@asu.edu}

\begin{abstract}

Lyman-$\alpha$ (\lya) is the most prominent ultraviolet emission line in low-mass stars, playing a crucial role in exoplanet atmospheric photochemistry, heating, and escape. However, interstellar medium (ISM) absorption typically obscures most of the \lya\ profile, requiring reconstructions that introduce systematic uncertainties. We present intrinsic \lya\ profiles for 12 high radial velocity K and M dwarfs, where Doppler shifting minimizes ISM contamination, allowing direct measurements of $\sim$50–95\% of the line flux. Our sample spans the K-to-M spectral transition, enabling us to constrain the dependence of self-reversals in \lya\ emission profiles on effective temperature ($T_{\rm eff}$). The depth of self-reversal, driven by non-local thermodynamic equilibrium (LTE) effects, decreases with decreasing $T_{\rm eff}$, with M dwarfs exhibiting little to none. Two stars, Ross 1044 and Ross 451, were observed over multiple days, revealing $\sim$20\% \lya\ variability confined to the line core - implying that studies relying on reconstructions may underestimate temporal variability. We find strong correlations between \lya\ flux, peak-to-trough ratio, and hydrogen departure coefficients with $T_{\rm eff}$, providing empirical constraints for stellar atmosphere models. A comparison of \lya\ flux in the habitable zone shows measured values for high radial velocity stars less than the reconstructed values for the rest of the sample, likely due to the older ages of the high-RV stars and/or overestimated reconstructed fluxes due to model deficiency (e.g., neglecting self-reversal). Our results establish an empirical foundation for \lya\ emission in K and M dwarfs, reducing uncertainties in reconstructions and improving models of stellar UV emission relevant to exoplanetary studies.

\end{abstract}



\section{Introduction} \label{sec:intro}

The \ion{H}{1} Lyman-$\alpha$ (\lya; $\lambda$1215.67 \AA) transition is the dominant ultraviolet (UV) emission feature in low-mass stars, with the intrinsic line flux estimated to contribute approximately 37\%–75\% of the total 1150–3100 \AA\ flux from most late-type stars \citep{France2013}. This emission plays a crucial role in stellar atmosphere modeling: the ionization balance of H I/H II in the outer stellar layers influences the overall atmospheric structure, and the line core is highly sensitive to the chromosphere and transition region, where non-radiative heating mechanisms remain poorly understood \citep{Short1997,Peacock2022}. Additionally, \lya\ radiation is a key driver of photodissociation in exoplanet atmospheres, breaking down molecules such as H$_2$O and CH$_4$ \citep[e.g.,][]{Rugheimer2015}. Since \lya\ controls atmospheric photochemistry, accurate estimates of its intrinsic flux are essential for assessing planetary habitability and interpreting atmospheric observations. Reconstructed \lya\ fluxes have been widely used to establish correlations with other spectral emission lines \citep{Linsky2013, Youngblood2017, Melbourne2020}, broadband UV photometry \citep{Shkolnik2014b}, and X-ray fluxes \citep{Linsky2020}. These correlations are often employed to assess the life-supporting potential of different stellar types \citep{Cuntz2016}, but systematic offsets may arise due to assumptions made in \lya\ reconstructions. 


Despite its importance, direct measurements of the intrinsic \lya\ line profile, flux, and variability are challenging due to severe absorption by neutral hydrogen in the interstellar medium (ISM) and contamination from geocoronal airglow, which affect over 99\% of observations. Standard approaches rely on reconstruction techniques that introduce uncertainties through assumptions about the intrinsic line shape and ISM properties along the line of sight \citep{Youngblood16, Wilson2022,Sandoval2023}. While significant effort goes into these reconstructions, they depend on assumptions about the shape of the line core and the structure of the ISM. \cite{Youngblood16} estimate that both could independently yield $\sim$30\% inaccuracies, while \cite{Sandoval2023} finds that in extreme cases, \lya\ flux estimates can be off by factors of 3--5. Additionally, \cite{Wilson2022} identified a 2$\times$ systematic uncertainty in the reconstruction process by comparing reconstructions of the M dwarf component of EG UMa, a tight white dwarf-M dwarf binary, performed at different orbital phases. Mg II self-reversed emission-line profiles provide some constraints on the \lya\ line core, but differences between the two lines - such as formation temperatures - are significant enough that Mg II cannot serve as a direct template for \lya\ reconstructions \citep{Youngblood2022}. 

Recent observations with the Space Telescope Imaging Spectrograph (STIS) onboard the Hubble Space Telescope (HST) have provided new insights by directly measuring intrinsic \lya\ profiles in six M and K-type stars with exceptionally large radial velocities ($|\text{RV}| > 85$ km s$^{-1}$), which Doppler shift the \lya\ emission away from these contaminating sources \citep{Schneider2019,Youngblood2022}. These observations reveal that self-reversal in the \lya\ line core is common among low-mass stars and that the depth of self-reversal correlates with surface gravity, with lower-mass stars exhibiting weaker self-reversals. Previous \texttt{PHOENIX} stellar atmosphere models underpredicted the \lya\ core strength for these stars, highlighting the need for improved microphysics in the upper atmosphere models \citep{Peacock2022}. The self-reversal in \lya\ arises from non-local thermodynamic equilibrium (non-LTE) effects, as the line forms over extensive depths in the chromosphere and transition region where collisional processes become less dominant, allowing the line source function to depart from the Planck function. Modeling the \lya\ lines of these stars demonstrated the sensitivity of the core flux to departures from LTE in the $n=2$ state of H I at the boundary between the chromosphere and transition region. These departure coefficients indicate missing or incorrect opacities in the models, requiring adjustments to the minimum values set in these layers to reproduce observed \lya\ profiles. 

These findings have significant implications for reconstructing \lya\ emission in stars where direct measurements remain infeasible. Self-reversal in \lya\ profiles must be accounted for to avoid overestimations of intrinsic flux by as much as 60\%–100\% for G and K dwarfs and 40\%–170\% for M dwarfs \citep{Youngblood2022}. Moreover, improved \lya\ modeling directly translates to better predictions of the EUV spectrum \citep[e.g.,][]{Linsky2013,Peacock2019,Linsky2024}, which is crucial for understanding atmospheric escape and photochemistry in exoplanets.

The results from \cite{Youngblood2022} and \cite{Peacock2022} reaffirm that the depth of self-reversal in \lya\ increases with earlier spectral type; however, small sample sizes have limited precise characterization of how the line profile changes with uniform temperature steps of less than 1000 K. In this paper, we expand on these works with new HST STIS observations, adding six additional M and K stars for which $>$85\% of the intrinsic \lya\ flux can be directly measured, all with absolute radial velocities exceeding 100 km s$^{-1}$. These observations double the sample of low-mass stars for which \lya\ can be measured directly, uniformly sampling stars with $T_{\rm eff}$ from 3400 to 5500 K in steps of approximately 500 K. By increasing the number of high-RV targets with directly observed \lya\ profiles and refining stellar atmosphere models, this study enhances our ability to accurately predict stellar \lya\ fluxes. These improvements are essential for interpreting exoplanetary atmospheres and assessing habitability, particularly for planets orbiting low-mass stars where \lya\ dominates the high-energy radiation environment.

\begin{deluxetable*}{lccccc}[t!]
    \tablecaption{Observations \label{tab:obs}} 
    \tablehead{
    \colhead{Star} 
    & \colhead{Observation Date} 
    & \colhead{Orbits}
    & \colhead{Exposure Time (s)} 
    & \colhead{SNR$_{wing}$}
    & \colhead{SNR$_{core}$}}
    \startdata
    HD 64090  & 2022-Mar-26 & 1 & 1614\ & 5.0 & 9.2\\
    \hline
    HD 134439 & 2022-Jul-04 & 2 & 4324 & 5.8 & 13.9\\
    \hline
    HD 134440 & 2202-Jul-07 & 2 & 4324 & 4.6 & 12.8\\
    \hline
    Ross 451  & 2022-Oct-01 & 3 & 7882 & 4.2 & 6.3\\
    Ross 451  & 2022-Oct-02 & 3 & 7882 & \nodata & \nodata\\
    Ross 451  & 2022-Oct-03 & 3 & 7882& \nodata & \nodata\\
    \hline
    HIP 117795& 2022-Oct-15  & 3 & 7733& 4.6 & 11.8\\
    \hline
    L 802-6   & 2023-Jan-19 & 3 & 6840 & 5.4 & 6.2
    \enddata
    \tablecomments{Observations of targets from HST-GO-16646. All observations were taken with STIS/G140M centered at 1222 \AA. The right-most columns give the SNR per resolution element for the coadded spectrum in the wing (0.6 \AA\ from the line core, on the side furthest from the ISM absorption) and nearest to the core of the \lya\ line. For Ross 451, a single SNR value is listed because all visits were combined into one final spectrum.}
\end{deluxetable*}

\section{Observations and Reductions} \label{sec:obs}
We observed six low-mass K and M stars—HD 64090 (K0), HD 134439 (K1), HD 134440 (K2), HIP 117795 (K8), Ross 451 (M0), and L 802-6 (M3)—between March 26, 2022, and January 19, 2023, using HST/STIS (Table \ref{tab:obs}). These targets were selected based on their RVs, as our goal was to observe \lya\ emission lines shifted outside the regions strongly affected by ISM absorption and geocoronal emission. We identified low-mass stars ($T_{\rm eff} \leq 5500$ K) in the {\it Gaia} DR2 catalog\footnote{The {\it Gaia} DR2 catalog contains more than 7.2 million radial velocity measurements, and the Gaia EDR3 release did not include new radial velocity data at the time of the observing proposal submission.} with absolute RVs greater than 100 km s$^{-1}$. This threshold corresponds to a wavelength shift of $\approx$0.4 \AA, though the exact RV required to shift the \lya\ peak out of the contaminated wavelength range varies with the line of sight to each star. Different sight lines have ISM absorbers at different RVs, usually between $\pm$30 km s$^{-1}$ \citep{Redfield2008}.

These observations, obtained as part of HST GO Program 16646 (DOI: 10.17909/bx2r-na24), spanned a total of 22 orbits. For each target, we used the STIS/G140M grating with the 52\arcsec $\times$ 0\farcs1 slit, centered at 1222 \AA, to capture the \lya\ wavelength region.

During the data reduction process, we identified extraction issues in the standard pipeline for some targets (Ross 451, HIP 117795, and L 802-6), where the spectral trace was not properly located by the automated calSTIS pipeline. To correct this, we manually located the spectral traces in the flat-fielded images (flt files), then forced extraction of the 1D spectrum at this location with calSTIS. To maintain consistency, we manually identified the trace location for all spectra. In the manual extractions, we moved the background estimation regions to an offset of 10 pixels above and below the spectral trace with widths of 20 pixels. Calibration reference files were those current as of 2023 March 10. Additionally, a two-orbit visit for HIP 117795 on October 14 failed due to user error in the selection of the acquisition aperture (F25ND5 was used in place of the intended F25ND3). As a result, we only used the visit from October 15 for this star.

For targets with multiple observations, we coadded the background-subtracted spectra by first interpolating them onto a common wavelength grid with $\Delta\lambda$ = 0.05 \AA\ and then performing an exposure-time-weighted average. The signal-to-noise ratio (SNR) per resolution element in the wings and nearest to the core of \lya\ for each star is listed in Table \ref{tab:obs}, with values ranging from 4.2–5.8 in the wings and 6.2–13.9 near the core. The loss of two orbits for HIP 117795 did not significantly impact the data quality, as the SNR near the core remained relatively high at 11.8. Ross 451 and L 802-6 have the lowest core SNRs ($\sim$6), due to their lower radial velocities ($\sim$100–150 km s$^{-1}$), which provided a smaller Doppler shift away from contaminating sources compared to the other four stars ($> 230$ km s$^{-1}$). 

\begin{figure}[t!]
    \centering
    \includegraphics[width=0.95\textwidth]{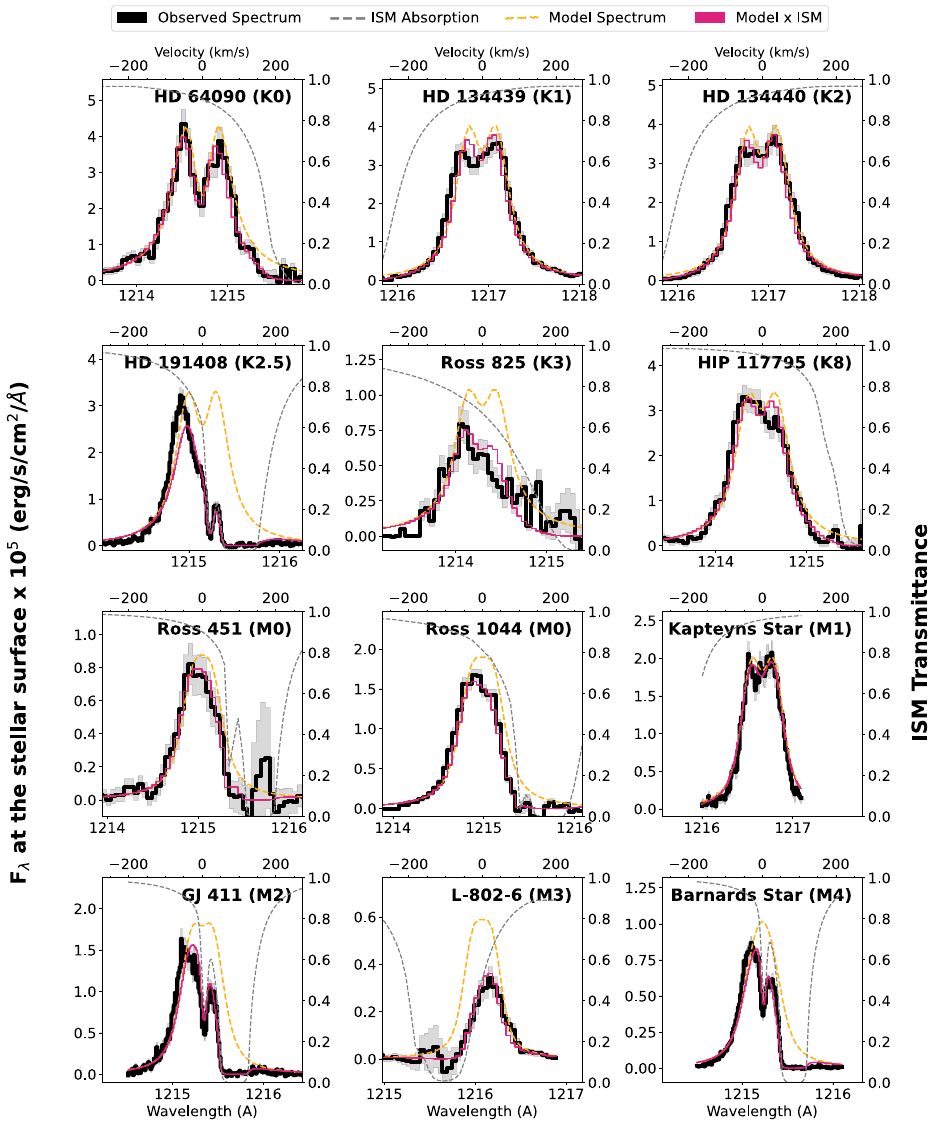}
    \caption{\lya\ profiles of low-mass K and M stars with high radial velocities ($|$RV$|$$> 84$ km s$^{-1}$) ordered by spectral type. HST/STIS observations are plotted in black, with associated errors shaded in gray. All stars were observed with the G140M grating, except for HD 191408, Kapteyn's Star, GJ 411, and Barnard's Star, which were observed with E140M. The ISM transmittance curve for each star is plotted as a gray dashed line. Intrinsic \texttt{PHOENIX} model profiles are plotted in yellow. The intrinsic \texttt{PHOENIX} model, after being multiplied by the ISM transmittance curve, is plotted in pink — this profile should match the observations. The yellow profile represents the intrinsic stellar emission before ISM absorption.}
    \label{fig:3x4}
\end{figure}

Figure \ref{fig:3x4} presents the coadded spectra for our six newly observed targets, alongside archival spectra of six additional low-mass stars with high radial velocities from \citet{Schneider2019} (Ross 825, Ross 1044) and \citet{Youngblood2022} (HD 191408, GL 411, Barnard's Star, and Kapteyn's Star). The Schneider et al. stars were observed with HST/STIS using the G140M grating, while the Youngblood et al. stars were observed with HST/STIS using the E140M echelle mode. These archival data are incorporated into our analysis to expand coverage in stellar parameter space and to strengthen our comparison across effective temperature. The full sample of 12 stars spans T$_{\rm eff}$ from 3300 to 5500 K in steps of 40 to 500 K (Table \ref{tab:stelprops}) and reveals nearly complete line profiles, with subtle core reversals that become more pronounced in hotter stars. Further details of additional components in Figure \ref{fig:3x4} are explained in Section \ref{sec:modeling}.

\begin{deluxetable*}{lccccccccccc}[t!]
    \rotate
    \tablecaption{Stellar Properties \label{tab:stelprops}} 
    \tablehead{
    \colhead{Star} 
    & \colhead{Sp. Type} 
    & \colhead{$T_{\rm eff}$} 
    & \colhead{log($g$)} 
    & \colhead{$[$Fe/H$]$}
    & \colhead{$M_\star$}    
    & \colhead{$R_\star$}  
    & \colhead{[Fe/H]}
    & \colhead{Distance} 
    & \colhead{$v$sin$i$}
    & \colhead{RV}
    & \colhead{Age}\\
    & 
    & 
    (K) & 
    (cm s$^{-2}$) &
    &
    ($M_\sun$) &
    ($R_\sun$) & 
    &
    (pc) &  
    (km s$^{-1}$)& 
    (km s$^{-1}$)&
    (Gyr)
    }
    \startdata
    HD 64090  & K0   & 5528 $\pm$ 35$^{a}$ & 4.62 $\pm$ 0.3 & -1.64 $\pm$ 0.15 & 0.9   & 0.62$\pm$0.04$^{m}$  &-1.64& 27.35 $\pm$ 0.02   & -0.3 - 8$^{y,m}$  & -234.1 $\pm$ 0.15  & 6.72$^{+5.6}_{-4.8}$ ($ac$)\\
    HD 134439 & K1   & 5065 $\pm$ 72$^{b}$ & 4.56 $\pm$ 0.1 & -1.46$\pm$ 0.1   & 0.8   & 0.60$\pm$0.01$^{m}$  &-1.46& 29.40 $\pm$ 0.01  & -0.87 - 7$^{n,o}$ & 310.3 $\pm$ 0.2   & 7.18 -- 9.9$^{ad,ae}$\\
    HD 134440 & K2   & 4796 $\pm$ 57$^{c}$ & 4.64 $\pm$ 0.1 & -1.41 $\pm$ 0.1  & 0.75  & 0.56$\pm$0.01$^{m}$  &-1.41& 29.39 $\pm$ 0.01  & -0.13 - 7$^{n,o}$ & 311.1 $\pm$ 0.2   & 6.12 -- 9.4$^{ad,ae}$\\
    HD 191408 & K2.5 & 4893 $\pm$ 49$^{d}$ & 4.6 $\pm$ 0.1  & -0.07 $\pm$ 0.03 & 0.82  & 0.74$\pm$0.17$^{n}$  &-0.07& 6.01 $\pm$ 0.01   & 3.02 - 3.97$^{p,q}$ & -129.3 $\pm$ 0.1 & 6.26 -- 11.24$^{ad,af}$\\
    Ross 825  & K3   & 4680 $\pm$ 170$^{e}$& 4.75 $\pm$ 0.2 & -1.28  $\pm$ 0.1 & 0.55  & 0.83$\pm$0.21$^{o}$  &-1.28& 98.26 $\pm$ 0.18 & \nodata           & -341.2 $\pm$ 0.5 & $>10^{ag}$\\
    HIP 117795& K8   & 4100 $\pm$ 150$^{f}$& 4.75 $\pm$ 0.25& -1.2 $\pm$ 0.3   & 0.62  & 0.52$\pm$0.02$^{m}$  &-1.2 & 26.73 $\pm$ 0.01 & \nodata           & -285.9$\pm$0.4  & \nodata\\
    Ross 451  & M0   & 3800 $\pm$ 25$^{g}$ & 4.73 $\pm$ 0.01& -0.96 $\pm$ 0.02 & 0.5   & 0.31$\pm$0.02$^{g}$  &-0.96& 24.74 $\pm$ 0.01 & \nodata           & -154.3 $\pm$ 0.3 & \nodata\\
    Ross 1044 & M0   & 3754 $\pm$ 95$^{h}$ & 4.99 $\pm$ 0.2 & -1.01 $\pm$ 0.21 & 0.3   & 0.38$\pm$0.03$^{h}$  &-1.01& 26.71 $\pm$ 0.02 & \nodata           & -168.6 $\pm$ 0.3 & $>10^{ah}$\\
    Kapteyn's Star & M1 & 3650 $\pm$ 80$^{i}$ & 4.96 $\pm$0.13 & -0.5 $\pm$ 0.3 & 0.28  & 0.28$\pm$0.01$^{p}$  &-0.5 & 3.93 $\pm$ 0.01  & -4.2 - 9.15$^{o,r}$ & 245.1 $\pm$ 0.1 & 11.5$^{+0.5}_{-1.5}$ ($ai$)\\
    GJ 411    & M2   & 3530 $\pm$ 100$^{j}$& 4.5 $\pm$ 0.01 & -0.2 $\pm$ 0.1   & 0.44  & 0.41$\pm$0.03$^{g}$  &-0.2 & 2.55 $\pm$ 0.01  & -3.25 - 6.5$^{o,s}$ & -85.1 $\pm$ 0.1 &$>5^{ag}$\\
    L 802-6   & M3   & 3384 $\pm$ 61$^{k}$& 5.08 $\pm$ 0.1 & -0.47 $\pm$ 0.1  & 0.4   & 0.25$\pm$0.11$^{m}$  &-0.47& 11.50 $\pm$ 0.01 & 1.70$\pm$0.38$^{t}$ & 101.4 $\pm$ 0.4 & 3.04$\pm$3.09$^{aj}$\\
    Barnard's Star & M4 & 3332 $\pm$ 50$^{l}$& 5.13 $\pm$ 0.01& -0.29 $\pm$ 0.2 & 0.23  & 0.187$\pm$0.01$^{q}$ &-0.29& 1.83 $\pm$ 0.01  & 3.1 $\pm$ 1.2$^{u}$ & -110.5 $\pm$ 0.1 & 7--10$^{ak}$
    \enddata
    \tablecomments{For each star, the $T_{\rm eff}$, log($g$), and $[$Fe/H$]$ were obtained from a single reference and validated against visible/infrared photometry from the Vizier Photometric Viewer\footnote{\url{http://vizier.unistra.fr/vizier/sed/}}. All distance and RV measurements are from \cite{Gaia2021} and estimated masses from \cite{Pecaut2013}. Uncertainties on $T_{\rm eff}$, log($g$), and $[$Fe/H$]$ were estimated from inter-quartiles of values listed on \url{simbad.u-strasbg.fr} if the listed reference did not provide them.}
    \tablerefs{
    (a) \cite{Luck2017};
    (b) \cite{Bensby2014};
    (c) \cite{Perrin1975};
    (d) \cite{Valenti2005};
    (e) \cite{Stassun2018};
    (f) This Work;
    (g) \cite{Kesseli2019};
    (h) \cite{Newton2015};
    (i) \cite{Anglada2014};
    (j) \cite{Lepine2013};
    (k) \cite{Gaidos2014};
    (l) \cite{Rosenthal2021};
    (m) \cite{Gaia2021};
    (n) \cite{Zacharias2012};
    (o) \cite{Glebocki2005};
    (p) \cite{Youngblood2022};
    (q) \cite{Peacock2022};
    (r) \cite{Houdebine2010};
    (s) \cite{Jonsson2020};
    (t) \cite{Das2025};
    (u) \cite{Fouque2018};
    (y) \cite{Stanford2020};
    (ac) \cite{Isaacson2010};
    (ad) \cite{Isaacson2010};
    (ae) \cite{Reggiani2018};
    (af) \cite{Harada2024};
    (ag) \cite{Gagne2018};
    (ah) \cite{Schneider2019};
    (ai) \cite{Wylie-de2010};
    (aj) \cite{Maldonado2020};
    (ak) \cite{ribas2018}
    }
\end{deluxetable*}

\section{Modeling Intrinsic \lya\ Profiles with \texttt{PHOENIX}}\label{sec:modeling}

The high radial velocities of the 12 stars shift the \lya\ line by 0.34 to 1.38 \AA, revealing 47\% to 95\% of the intrinsic profiles (Table \ref{tab:lya_flux}) that would otherwise be mostly obscured by the ISM. To ensure a fair comparison between models and observations, we applied the appropriate ISM absorption for each target’s line of sight. We computed ISM transmittance curves for each target by performing a reconstruction following methods from \cite{Youngblood2022}. Briefly, we forward-modeled the observed spectrum with an intrinsic stellar emission profile and an ISM absorption profile using \texttt{lyapy} \citep{lyapy}. The product of the two profiles is convolved with the instrument line spread function. We assume a self-reversed Voigt profile for the stellar emission and a single absorption component for the ISM. The optical depth of the \ion{H}{1} and \ion{D}{1} lines are modeled as Voigt profiles. For the subsequent analysis, we solely retain the ISM absorption profiles.

To reproduce the intrinsic \lya\ profiles for the complete set of 12 stars,
we computed models using the \texttt{\texttt{PHOENIX}} atmosphere code \citep{Hauschildt1993, Hauschildt2006, Baron2007}, following the methodology of \cite{Peacock2022}. For each modeled star, we constructed photospheric structures based on literature values for effective temperature ($T_{\rm eff}$), mass (M$_\star$), surface gravity (log($g$)), and metallicity ([Fe/H]) (Table \ref{tab:stelprops}). To these photospheres, we added chromospheric and transition region layers modeled as linear temperature rises as a function of log(column mass), reaching a maximum temperature of 2 $\times$ 10$^5$ K. This upper limit exceeds the temperature range in which \lya\ forms ($\approx$2$\times$10$^3$--8$\times$10$^4$ K) and is consistent with our previous analysis. The temperature at the top of each chromosphere, where hydrogen becomes fully ionized and the atmosphere becomes thermally unstable, ranges from 7000 to 8000 K. In \cite{Peacock2022}, we confirmed that assuming a linear temperature rise with log(column mass) when modeling the stellar chromosphere yields spectra that best match UV observations. Additionally, we found that smoothing the temperature structure at the chromosphere-transition region boundary has a negligible effect on the computed UV spectrum, including the \lya\ profile.

The \lya\ profiles were computed assuming Voigt functions. We included line blanketing in the background opacities, computed \lya\ with partial frequency redistribution, and incorporated the same robust set of species computed in non-local thermodynamic equilibrium (non-LTE) as in \cite{Peacock2022}. We used a microturbulent velocity of 2 km s$^{-1}$ in the photosphere and a velocity gradient in the chromosphere and transition region set as a fraction of the local sound speed (0.35$\times$ $v_{sound}$), with a maximum velocity capped at 10 km s$^{-1}$.

For each star, we computed a grid of 72 distinct upper atmosphere models, varying the location and thickness of both the chromosphere and transition region using three free parameters: (1) the base of the chromosphere, (2) the top of the chromosphere, and (3) the temperature gradient in the transition region. A key difference from \cite{Peacock2022} is that not all stars in this study have NUV observations to provide additional empirical constraints on the upper atmospheric structure. Four stars (HD~191408, Kapteyn’s Star, GJ~411, and Barnard’s Star) have archival \textit{HST}/STIS E230H spectra (with some also having additional gratings), while six others (Ross~825, Ross~1044, HD~64090, HIP~117795, Ross~451, and L802-6) have GALEX NUV photometry. Two additional stars (HD~134439 and HD~134440) are listed as having STIS G230LB observations, but the data are not publicly available. Because of this heterogeneity in NUV coverage and data quality across the sample, we chose to fit only the Ly$\alpha$ line in order to apply a consistent modeling approach to all 12 stars. To compare the models with observations, we applied the radial velocity shifts, multiplied by the appropriate ISM transmittance curves, scaled the flux by $R_\star^2/dist^2$, and then convolved the model spectra to the observational resolution. As in \cite{Peacock2022}, our initial model setup yielded spectra that successfully reproduced the observed \lya\ line widths, but severely underestimated the line core (see Appendix). 

The departure coefficients, which represent the ratio of non-LTE to LTE number densities, serve as proxies for the population of each energy level and are essential for calculating emissivity and absorption coefficients. The \lya\ line is produced by electron transitions from the n=2 state to the ground state (n=1) in hydrogen. In \cite{Peacock2022}, we showed that in our original model setup, the departure coefficient for the n=2 state drops by several orders of magnitude at the chromosphere-transition region boundary (also shown in Figures \ref{fig:a1}, \ref{fig:a2}, and \ref{fig:a3} in the Appendix). This sharp decrease indicates a severe underprediction of the hydrogen n=2 population in this region, whereas observations suggest it should be much closer to the LTE population. A key finding of that study was that the strength of the self-reversal in the \lya\ line core is highly sensitive to the minimum value set in the layers around the chromosphere-transition region boundary. We found that manually increasing the n=2 ratio across the upper chromosphere, where previous downturns occurred, enhances the flux in the \lya\ core without affecting the Lyman continuum. 

Building on this, for the present study, from each star's 72-model grid, we identified chromospheric structures that fit the uncontaminated wings of the observed \lya\ profiles within a reduced chi-squared threshold of $\Delta\chi_{\nu}^2\leq4$ from the minimum (corresponding to 1$\sigma$ uncertainty, given four three parameters). From these new subsets of models, we then generated updated grids by varying the minimum n=2 level population of H I at the chromosphere-transition region boundary, ranging from 0.3 to the maximum value in the chromosphere (typically $\sim$3) in steps of 0.1. As a result, the new grids contain sets of \lya\ profiles with similar line widths but varying reversal depths\footnote{Similar to findings in \cite{Peacock2022}, adjusting this minimum has negligible effect on the rest of the computed spectrum, with the flux across EUV wavelengths (100 -- 912 \AA) and the H$\alpha$ line changing by $<$ 1\%. This lack of change does not imply that \lya\ should not correlate with either, but rather that they are less sensitive to changes in the $n=2$ population of H I in these layers.}. 

We again applied the radial velocity shifts, multiplied by the appropriate ISM transmittance curves, and convolved the model spectra to the observational resolution. This time, instead of selecting models based solely on the uncontaminated wings, we identified the model with a $\chi_{\nu}^2$ value closest to 1 across the full line width. We present these best-fitting models in Figure \ref{fig:3x4}, where both the intrinsic model profiles and those modified by ISM transmittance are plotted against the observations (Appendix Figures \ref{fig:a1}, \ref{fig:a2}, and \ref{fig:a3} show the adjustments made to the  minimum n=2 level population of H I at the chromosphere-transition region boundary). The intrinsic profiles confirm moderate-to-no central reversals in K and M stars, with the deepest reversal appearing in the hottest star in our sample ($\sim$5500 K) and no reversal in the coolest ($\sim$3300 K). We analyze trends with these intrinsic \lya\ profiles in Section \ref{sec:analysis}.

\begin{figure}[t!]
    \centering
    \includegraphics[width=0.9\textwidth]{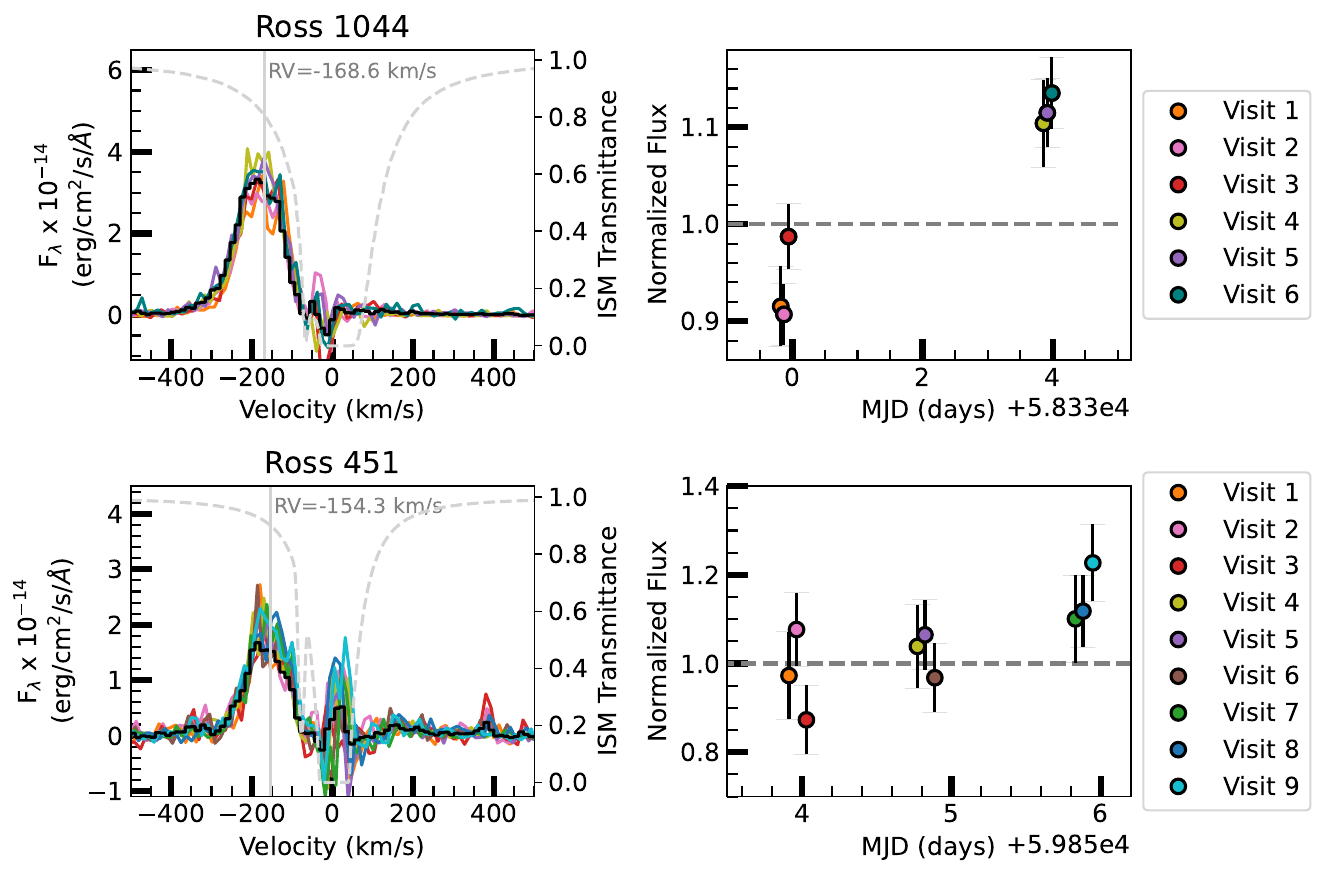}
    \caption{Of the 12 high RV stars with \lya\ measurements, two have observations taken over multiple days and both exhibit variability on the order of 20\% over the 3-4 day observing windows. \textit{Left}: HST/STIS \lya\ spectra of Ross 1044 (top) and Ross 451 (bottom) over multiple visits (various colors) compared to the final coadded spectrum in black. The ISM transmittance curves are plotted as gray dashed lines. \textit{Right}: \lya\ flux variability of Ross 1044 (top, adapted from \citealt{Schneider2019}) and Ross 451 (bottom), as measured over the observed spectrum from $\pm$0.5 \AA\ from the line center. Each set of points are normalized to the flux of the final coadded spectrum.}
    \label{fig:variability}
\end{figure}

\begin{figure}[t!]
\centering
\includegraphics[width=0.44\textwidth]{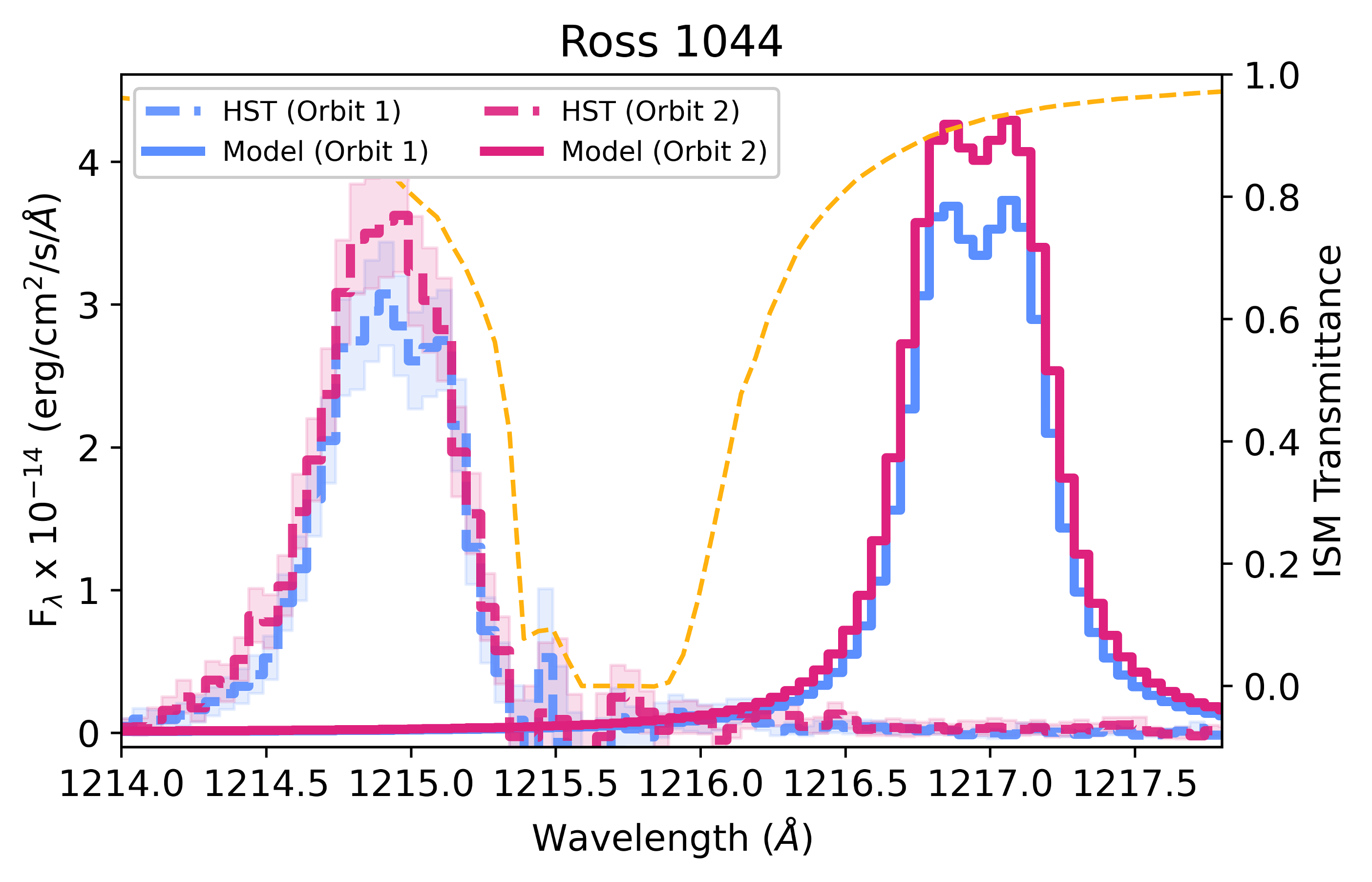}
\includegraphics[width=0.45\textwidth]{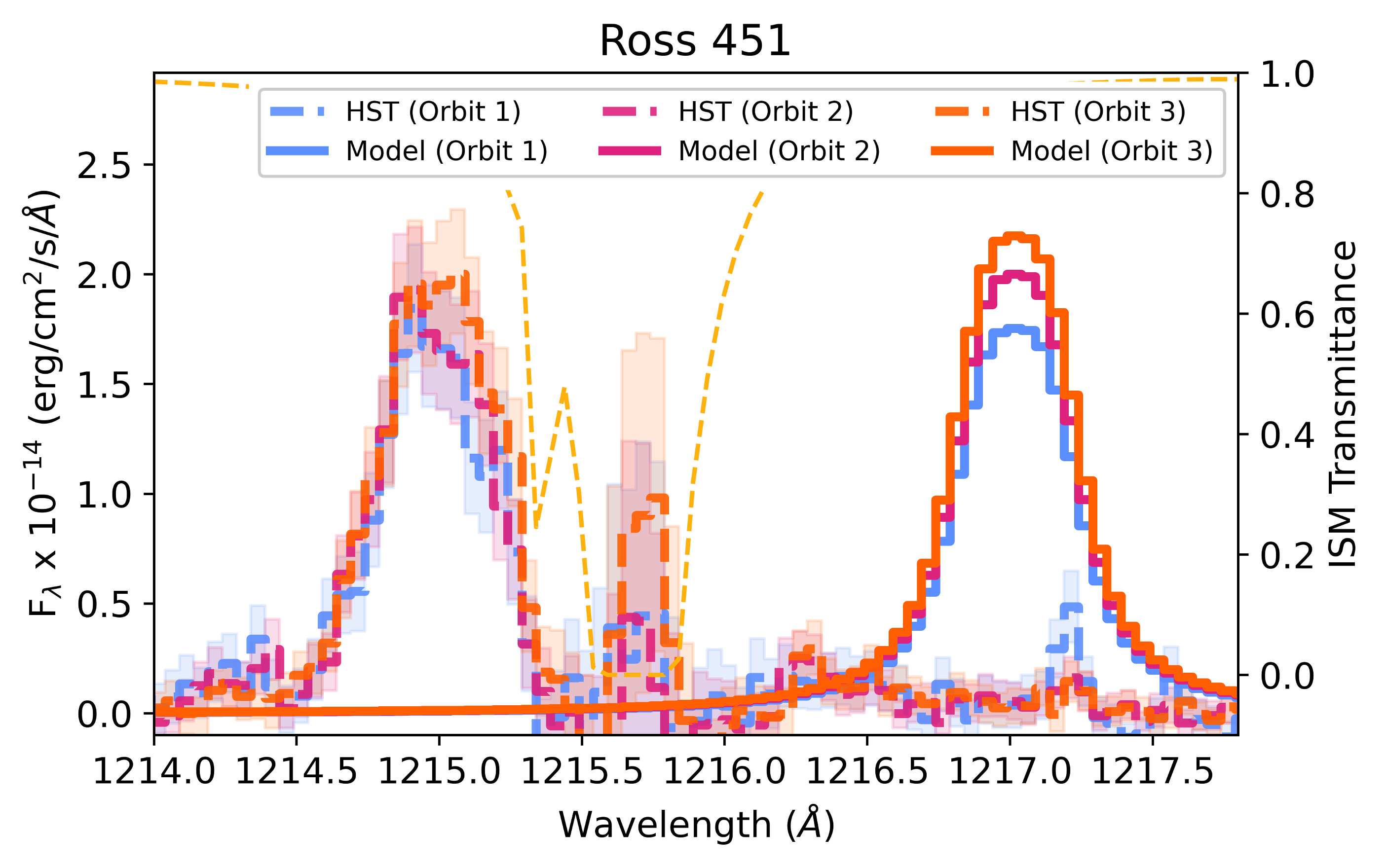}
\caption{Measured (dashed) and modeled (solid; and shifted in wavelength for visual clarity) \lya\ profiles for Ross 1044 (\textit{left}) and Ross 451 (\textit{right}) from the multiple HST orbits that exhibited variable behavior. The modeled profiles show the stellar emission in the absence of ISM absorption. For Ross 1044, there is a 5\% difference in self-reversal depth as measured via peak-to-trough ratio; there is no difference in peak-to-trough ratio for Ross 451. The ISM transmittance curve is plotted as the yellow dashed line.}
\label{fig:variable_r1044}
\end{figure}

\subsection{Variable Stars}\label{sec:variability}

For most stars, \lya\ variability can originate from three primary sources: 1) intrinsic stellar processes (e.g., rotation or magnetic activity), 2) escaping hydrogen from a transiting planet undergoing mass-loss, and 3) instrumental effects. For HST/STIS, there exists a breathing effect caused by thermal cycling as the spacecraft moves between orbital day and night that can introduce small flux variations within ($<$10\%) and between (1--3\%) orbits \citep{Kimble1998,Brown2001,Ben-Jaffel2007,Ehrenreich2012,Waalkes2019}.For the 52\arcsec $\times$ 0\farcs1 slit aperture used in the present observations, the photometric repeatability is approximately 8.8\% root mean square (RMS), consistent with expectations for smaller STIS apertures subject to thermal breathing effects \citep{Bohlin1998}.

In the Sun, \lya\ flux varies by 1-30\% during minutes-long flares \citep{Milligan2020,Greatorex2023}, 25–40\% over the 27-day solar rotation cycle and by 80–155\% over its 11-year solar cycle \citep{bossy1981, lean1987}. The core of the line shows the most variability, the wings of the profile remain relatively stable. At solar maximum, the increased emission partially fills in the self-reversal, leading to a lower peak-to-trough ratio \citep[e.g.,][]{Curdt2010,Kowalska2018, Gunar2020}. 

In low-mass stars, which are typically more active than solar-type stars, similar variability has been observed in other UV spectral lines \citep[e.g.,][]{Loyd2014,duvvuri2023}, with measured variability ranging from 1\% to 41\% over short timescales (minutes to hours). Short-term variations are caused by flares, fluctuations and stochastic variability, intermediate-term variations (weeks to months) result from stellar rotation, active region evolution, and episodes of major activity, while long-term variations (years) arise from stellar cycles and the evolution of active regions.

A common characteristic of high-RV stars is that they are typically old (Table \ref{tab:stelprops}), having had sufficient time for dynamical interactions within the Galaxy to increase their motion and shift them to high radial velocities \citep[e.g.,][]{Wielen1977,Nordstrom2004}. Older stars generally exhibit lower levels of UV activity than their younger counterparts; however, M and K dwarfs are still capable of regular flare activity, even in seemingly ``inactive" stars \citep{Loyd2018b,france2020}. While \citealt{Loyd2018b} found that \lya\ typically shows little response to flares in low-mass stars, those observations lacked information on the line core. 

In contrast, we find evidence of detectable \lya\ variability in two old, high-RV stars on timescales that could be related to flare activity or stellar modulation (Figure \ref{fig:variability}). Specifically, we detect \lya\ variability in one of our new targets, Ross 451, an M0V star. \citealt{Schneider2019} previously reported similar variability in the other M0V star in this sample, Ross 1044. Both stars exhibit $\approx$20\% flux variations across multiple visits, primarily in the core of the emission line. As these variations are much larger than the 8.8\% flux changes associated with STIS thermal breathing for this aperture size \citep{Bohlin1998}, and are confined to the line core rather than producing wavelength-independent fluctuations across the entire profile, we interpret them as likely intrinsic to the star rather than instrumental in origin. Among the 12 high-RV stars with \lya\ measurements, these are the only two observed over several days, and both display similar behavior on comparable timescales. The other stars with multiple orbits had their observations occur on a single day within a $<$ 3 hour period (excluding the failed visit of HIP 117795 on Oct 14).

When modeling the observations for these two stars individually by orbit (coadding the 3 visits per orbit), we find that for Ross 451, the line strength varies between orbits; however, each observation is best fit by a line profile without self-reversal, with a consistent peak-to-trough ratio of 1 across all three orbits. In contrast, for Ross 1044, the two observed orbits are best fit by different models, each yielding slightly different peak-to-trough ratios of 1.06 and 1.11 (Figure \ref{fig:variable_r1044}). This behavior is reminiscent of the solar Lyman-alpha profile, which also exhibits variations in self-reversal depth and peak-to-trough ratio over the solar cycle due to changes in chromospheric activity. The observed variability in Ross 1044 suggests that localized enhancements in chromospheric emission, such as flares or active regions, may temporarily reduce the depth of the self-reversal. Additionally, rotational modulation could contribute, as active regions or plage rotating in and out of view may alter the observed profile over time, similar to how solar Lyman-alpha emission varies with the passage of active regions across the solar disk.

Ross 451 and Ross 1044 are both M0 dwarfs with similar radial velocities (–168.6 and –154.3 km/s), and both show comparable levels of \lya\ variability. While the alignment in velocity and variability might suggest a common cause, it appears to be a coincidence. In principle, variability could be due to ISM effects or instrumental limitations (e.g., the use of a narrow 0\farcs1 slit). However, given that both stars are largely clear of the ISM (the transmittance at the line core is $\approx$80\%) and that the structure of an intervening ISM cloud is unlikely to vary on timescales of a few days, the observed changes are unlikely to be caused by ISM variability. Instrumental effects, such as potential slit losses associated with the narrow 0\farcs1 aperture, cannot be entirely ruled out, but the variability is strongly wavelength-dependent: when the flux is computed in 0.1 \AA\ bins across the line profile, the variability ranges from $\approx$10\% to 60\%. This wavelength dependence, combined with the temporal behavior, points to an origin intrinsic to the star. \citet{Schneider2019} previously suggested that the \lya\ variability in Ross 1044 may arise from rotational modulation or stellar flares, an interpretation that aligns with our results.


\section{Analysis}\label{sec:analysis}

\begin{deluxetable*}{lccccc}[t!]
    \tablecaption{\lya\ Line Properties \label{tab:lya_flux}} 
    \tablehead{
    \colhead{Star} 
     & \colhead{\% of line}
     & \colhead{Peak-to-trough}
     & \colhead{log $N$(H I)}
    & \colhead{F$_{\lya, surface}$} 
     & \colhead{F$_{\lya, HZ}$} \\
     &
     exposed&
     Ratio&
     (cm$^{-2}$)&
     (erg cm$^{-2}$ s$^{-1}$) &
     (erg cm$^{-2}$ s$^{-1}$)
    }
    \startdata
    HD 64090  & 83\%    & 1.88 &18.46$^{+0.15}_{-0.17}$ & 2.24 $\pm$ 0.07 $\times$10$^5$ & 5.80 $\pm$ 1.75\\
    HD 134439 & 89\%    & 1.21 & 18.83$^{+0.09}_{-0.11}$ & 2.51 $\pm$ 0.07 $\times$10$^5$& 8.74 $\pm$ 2.57\\
    HD 134440  & 88\%   & 1.28 & 18.51$^{+0.26}_{-0.46}$ & 1.48 $\pm$ 0.07 $\times$10$^5$& 5.80 $\pm$ 1.68\\
    HD 191408 & 49\%    & 1.28 & 18.28$\pm$0.02$^a$ & 2.88 $\pm$ 0.05 $\times$10$^5$& 12.14 $\pm$ 3.50\\
    Ross 825  & 69\%    & 1.14 & 18.80$^b$ & 4.90 $\pm$ 0.21 $\times$10$^4$& 2.25 $\pm$ 0.64\\
    HIP 117795 & 92\%   & 1.14 &18.13$^{+0.53}_{-0.42}$& 2.40 $\pm$ 0.08 $\times$10$^5$& 17.88 $\pm$ 4.87\\
    Ross 451  & 84\%    & 1.00 & 18.41$^{+0.28}_{-0.48}$ & 3.24 $\pm$ 0.51 $\times$10$^4$& 3.22 $\pm$ 0.85\\
    Ross 1044 & 85\%    & 1.00 & 18.86$^b$ & 7.94 $\pm$ 0.08 $\times$10$^4$& 8.27 $\pm$ 2.19\\
    Kapteyn's Star&95\% & 1.10 & 17.98$^{+0.36}_{-0.32}$$^a$& 1.05 $\pm$ 0.05 $\times$10$^5$& 12.14 $\pm$ 3.18\\
    GJ 411    & 60\%    & 1.02 & 17.84$\pm$0.03$^a$ & 7.24 $\pm$ 0.06 $\times$10$^4$& 9.55 $\pm$ 2.48\\
    L 802-6   & 47\%    & 1.00 & 18.39$^{+0.25}_{-0.33}$ & 2.75 $\pm$ 0.17 $\times$10$^4$& 4.28 $\pm$ 1.10\\
    Barnard's Star&64\% & 1.00 & 17.72$\pm$0.03$^a$ & 3.62 $\pm$ 0.04 $\times$10$^4$& 5.99 $\pm$ 1.53
    \enddata
    \tablecomments{Fraction of \lya\ flux exposed is calculated as $\frac{(intrinsic\ model \times ISM)}{intrinsic\ model}$. \lya\ fluxes are derived from the intrinsic models by integrating over $\pm$ 0.75 \AA\ from line center, with uncertainties propagated from the observational values. Habitable zone fluxes are computed following \cite{Kopparapu2014}, assuming a 1 Earth-mass planet located at the midpoint between the Recent Venus and Early Mars limits.}
    \tablerefs{(a) \citealt{Youngblood2022} ; (b) \citealt{Schneider2019}}
\end{deluxetable*}

\cite{Ayres1979}, \cite{Linsky1980}, \cite{Youngblood2022} and \cite{Taylor2024} have connected chromospheric emission line properties as a function of chromospheric heating, $T_{\rm eff}$, surface gravity, and elemental abundance. With the intrinsic profiles presented in this paper, we assess whether surface \lya\ line flux and the self-reversal depth (quantified as the peak-to-trough ratio) are also connected.

\subsection{Connections Between \lya\ Flux and Stellar Parameters}

\begin{figure}
    \centering
    \includegraphics[width=0.75\linewidth]{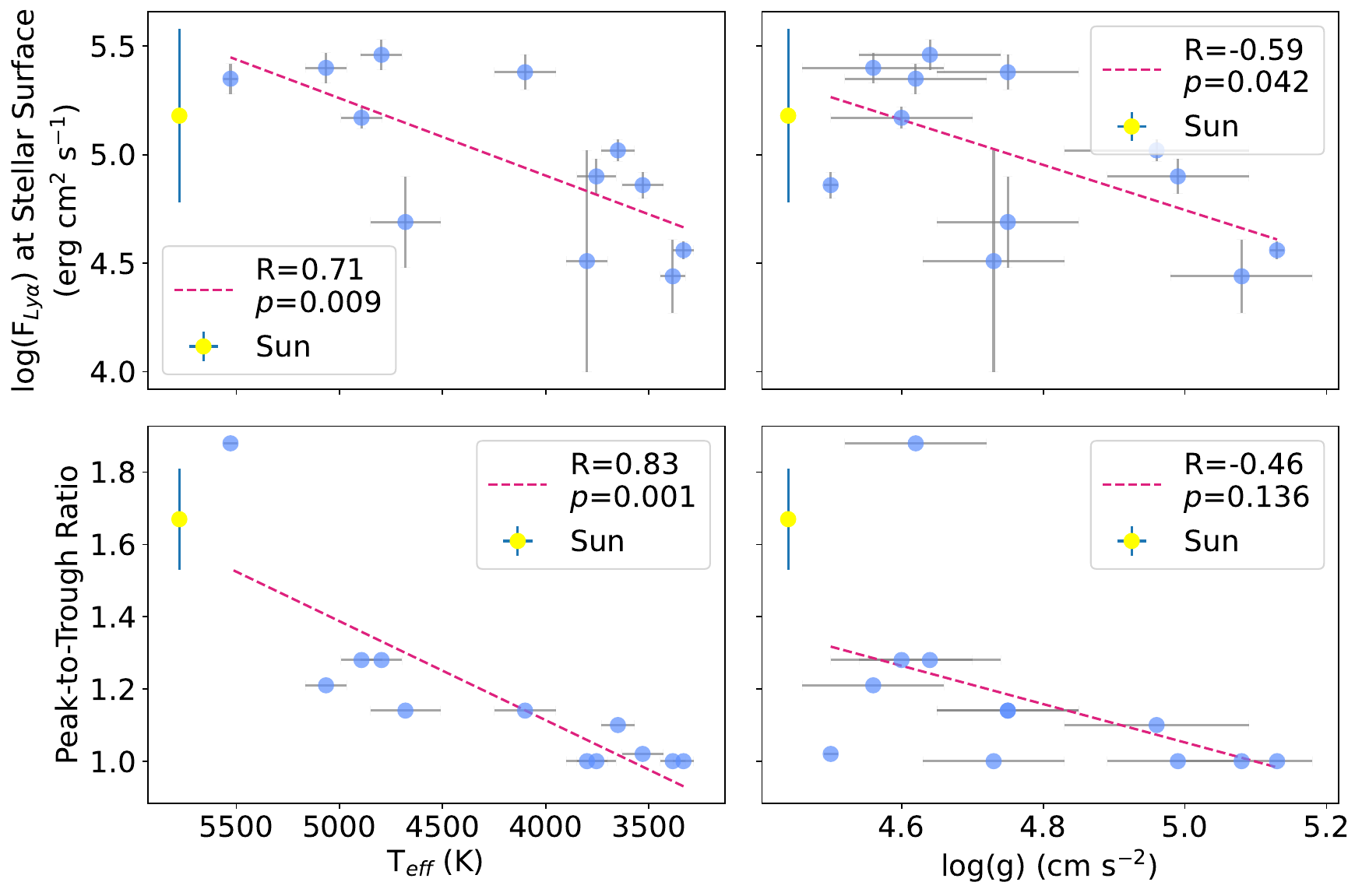}
    \caption{The logarithm of $\lya$ flux (log(F$_{\lya}$)) at the stellar surface (\textit{top row}) and peak-to-trough ratio of the intrinsic $\lya$ profile (\textit{bottom row}) versus $T_{\rm eff}$ (\textit{left column}) and the logarithm of surface gravity (\textit{right column}). Linear regression fits are plotted as pink dashed lines and the Sun is plotted as a yellow circle (the Sun is excluded from the fits). We find strong trends (R$>$0.7, \textit{p} $<$0.01) between $T_{\rm eff}$ and both log(F$_{\lya}$) and peak-to-trough ratio. We note that the K0 star HD 64090 (5528~K) exhibits a markedly deeper self-reversal than the rest of the sample, including the Sun, and appears as an outlier in the peak-to-trough ratio plots. This pronounced reversal is likely attributable to the star’s old age and correspondingly low activity level, consistent with the behavior observed in the Sun, whose core reversal deepens during periods of low activity and becomes more filled in when activity is higher. }
    \label{fig:correlations_p}
\end{figure}

In Figure \ref{fig:correlations_p}, we examine trends in \lya\ profile properties (Table \ref{tab:lya_flux}) as functions of $T_{\rm eff}$ and surface gravity. While instrumental broadening can reduce the measured peak-to-trough ratio, our calculations are based on intrinsic models treated consistently, ensuring that trend identification remains unaffected. However, we note that the method for quantifying the Sun’s values, taken from \cite{Youngblood2022}, differ from the rest of our sample and are included for context rather than in the linear regression fits.

We find strong correlations (R $>$ 0.7, \textit{p} $<$ 0.01) between both \lya\ surface flux and peak-to-trough ratio with $T_{\rm eff}$, indicating that hotter stars emit more \lya\ surface flux. A 5500 K star emits approximately ten times more \lya\ flux at its surface than a 3500 K star. Consistent with previous studies, we observe deeper self-reversal in earlier spectral types, with all K stars exhibiting some degree of reversal, whereas four of the six M stars suggest no self-reversal, and the other two only slight. 

There is overlap between three M stars analyzed in \cite{Youngblood2022} and those in our sample: Kapteyn’s Star, GJ 411, and Barnard’s Star. \cite{Youngblood2022} used reconstruction methods that tend to favor very slight reversal depths for M stars (shown in Appendix Figure \ref{fig:a3}), reporting peak-to-trough ratios of 1.11 $\pm$ 0.04 for Kapteyn’s Star, 1.10$^{+0.05}_{-0.04}$ for GJ 411, and 1.03 $\pm$ 0.02 for Barnard’s Star. In our analysis, we examine six M stars in total. Of the three M stars not in \cite{Youngblood2022}, none exhibit self-reversals. For the three overlapping stars, we find no reversal in Barnard’s Star (1.0), a slightly smaller reversal in GJ 411 (1.03), and the same value for Kapteyn’s Star (1.11). These results further reinforce the trend that self-reversal is minimal or absent in M stars, with only slight variations depending on the reconstruction method used.

In line with the influence of reconstruction method choices and the limitations of small-number statistics, \cite{Youngblood2022} found a clear correlation between peak-to-trough ratio and log(g), with lower log(g) corresponding to deeper self-reversal. We see a similar trend, though with a weaker linear regression fit. Our larger sample (12 stars vs. 5, or 13 vs. 6 including the Sun) contributes to this difference, along with discrepancies such as \cite{Youngblood2022} finding a stronger reversal depth for HD 191408 and including 82 Eri (G8V), which we excluded due to spectral type.

\subsection{Identifying Trends with Atmospheric Structure}\label{sec:atmo-trends}

\begin{figure}
    \centering
    \includegraphics[width=0.75\linewidth]{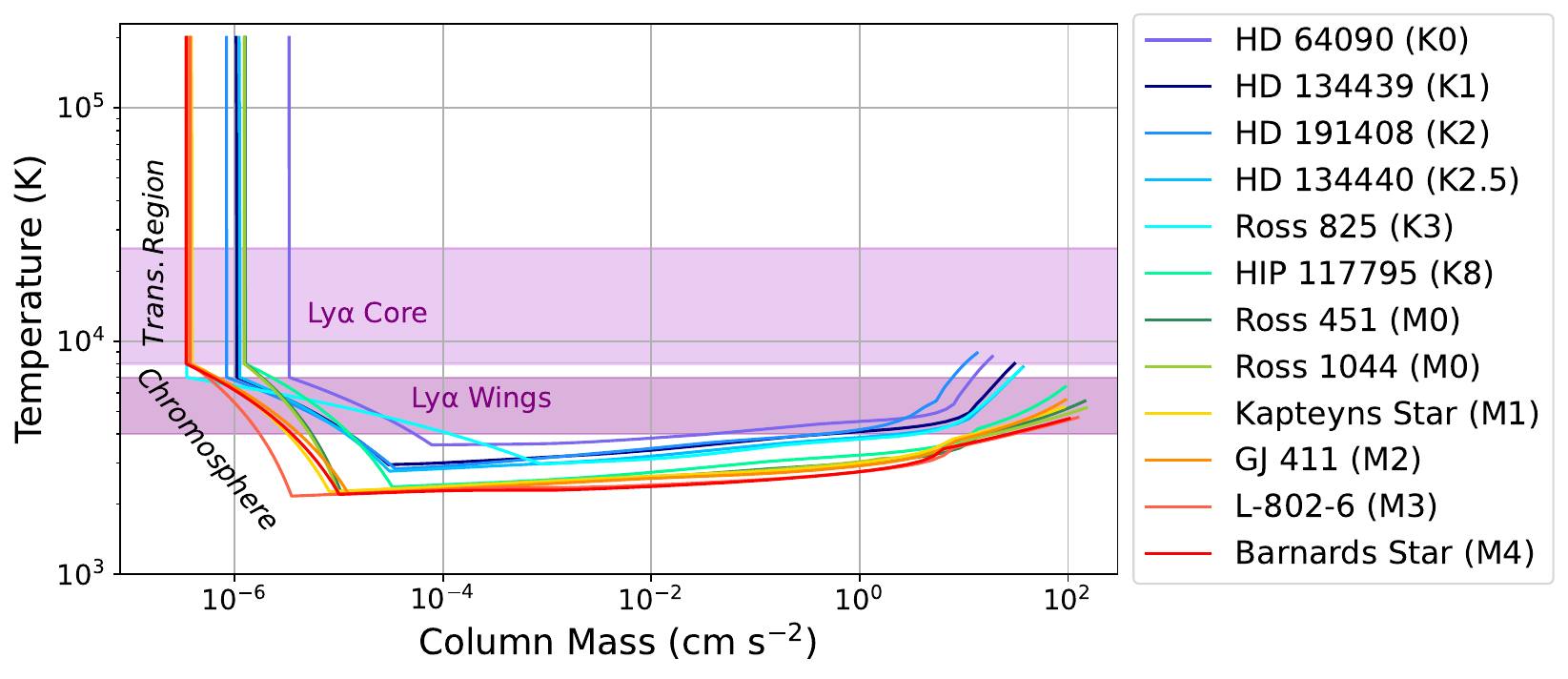}
    \caption{Temperature-column mass structures for the models of each star. We shade the approximate temperature ranges over which the core and wings of \lya\ forms in purple ($\approx$8000--25,000 K for the core, and $\approx$4000--7500 K for the wings), the actual range of temperatures varies by star and is determined by where the optical depth, $\tau(\lambda)$, equals to one.}
    \label{fig:structure}
\end{figure}

The stark difference in self-reversal depths between K and M stars suggests a fundamental shift in the structure and dynamics of their upper atmospheres. This may be linked to differences in chromospheric activity, turbulence, or radiative transfer effects. To investigate this further, we examine trends in the model chromospheric structures that best reproduce the observed \lya\ profiles. Figure \ref{fig:structure} shows the temperature-column mass structures for the 12 stars, with shaded regions highlighting the approximate temperature ranges where the wings and core of \lya\ forms. A clear pattern emerges: in K stars, the chromosphere tends to start deeper in the atmosphere, at column masses around 10$^{-4.5}$ to 10$^{-4}$ cm s$^{-1}$, and extends over a larger range ($\Delta$cmass = 10$^{1.5}$ cm s$^{-1}$). M stars, on the other hand, have more compressed chromospheres that initiate at lower column masses, between column masses of 10$^{-5.5}$ to 10$^{-5}$ cm s$^{-1}$, with thicknesses of $\Delta$cmass = 10$^{1}$ - 10$^{1.5}$ cm s$^{-1}$. Ross 825 stands out as an outlier, preferring a much broader chromosphere ($\Delta$cmass =10$^{3.5}$ cm s$^{-1}$) that begins at 10$^{-3}$ cm s$^{-1}$. 

In the present work, we focus exclusively on fitting \lya\ to maintain consistency across the sample and to identify general trends. It is important to note that since these models are constrained by a single spectral feature (the \lya\ line), they may not fully represent the broader UV spectrum. For example, Ross~825, Ross~1044, and Kapteyn’s Star were previously modeled in \citet{Peacock2022} using both NUV and \lya\ flux, resulting in notably different chromospheric structures. The chromospheric structure of Ross~825, which stands out as an outlier in the present \lya-only fits, illustrates the impact of including NUV constraints. As shown in \citet{Peacock2022}, when GALEX NUV photometry is simultaneously fit alongside \lya, the resulting model prefers a more compressed chromosphere, initiating at higher column mass (10$^{-5.5}$ to 10$^{-3.5}$~g~cm$^{-2}$) and aligning more closely with the other K stars in the sample. In contrast, Ross~1044 requires a significantly wider chromosphere when both \lya\ and NUV are fit, with the chromospheric temperature rise beginning as deep as 10$^{-2}$~g~cm$^{-2}$ and extending up to 10$^{-6.5}$~g~cm$^{-2}$, suggesting an extended heating region. Kapteyn's Star also shows a systematic shift: when NUV data are included, the entire chromosphere is shifted uniformly to lower column masses by $\Delta$cmass = 10$^{0.5}$~g~cm$^{-2}$, a trend consistent with the general behavior identified in this present study. These comparisons reinforce that while the \lya-only models capture key aspects of the upper atmospheric structure, simultaneous fits including NUV fluxes can shift or reshape the inferred chromospheres in systematic ways.

To assess the level of consistency between our \lya-only models and available NUV measurements, we spot-checked these same three stars by computing synthetic NUV fluxes from our best-fit models and comparing them to the observed values. The results show general agreement within a factor of a few: for Ross~825, the observed GALEX NUV flux is $(1.37 \pm 0.12) \times 10^{-12}$ erg s$^{-1}$ cm$^{-2}$, compared to a model flux of $1.80 \times 10^{-12}$ erg s$^{-1}$ cm$^{-2}$; for Kapteyn’s Star, the observed value is $(1.09 \pm 0.01) \times 10^{-12}$ erg s$^{-1}$ cm$^{-2}$ and the model predicts $2.94 \times 10^{-12}$ erg s$^{-1}$ cm$^{-2}$; and for Ross~1044, the observed flux is $(1.47 \pm 0.46) \times 10^{-13}$ erg s$^{-1}$ cm$^{-2}$ while the model yields $1.39 \times 10^{-13}$ erg s$^{-1}$ cm$^{-2}$. These comparisons suggest that our models constrained by \lya\ alone are broadly consistent with the available NUV data, despite not explicitly fitting to it.

The difference in chromospheric structures between K and M stars directly impacts the \lya\ flux. Although M star chromospheres are more compressed in column mass, they can still exhibit strong chromospheric emission because this compression leads to locally higher gas densities. This may seem counterintuitive, since their chromospheres begin at lower column masses than those of K stars. However, the key lies in the atmospheric structure: M dwarfs have smaller pressure scale heights, so a given change in column mass corresponds to a smaller change in geometrical height. As a result, material at a given column mass is located deeper in the atmosphere (at higher gas pressure and density) in M stars than in K stars. Thus, even though M dwarf chromospheres span a narrower range in column mass, the local densities in these layers can be higher, enhancing their chromospheric emission. Nevertheless, K stars tend to have deeper chromospheres overall, which leads to systematically higher \lya\ surface fluxes compared to M stars.

The transition region’s temperature gradient is extremely steep (so much so that it’s hard to distinguish in the plot) but no obvious trends emerge, with values spanning $\nabla$T$_{\rm TR}$ = 10$^{7.5}$-10$^9$ g$^{-1}$ cm$^{2}$. One key difference between K and M stars is the onset temperature of the transition region: around 7000 K for K stars and 8000 K for M stars. This is set by the point where hydrogen becomes fully ionized, triggering thermal instability in the transition region. This is the same part of the atmosphere where we imposed minimum values for n$_{NLTE}$ /n$_{LTE}$ in the n=2 state of H~I in order to reproduce the observations, specifically the flux in the \lya\ core.

\begin{figure}
    \centering
    \includegraphics[width=0.75\linewidth]{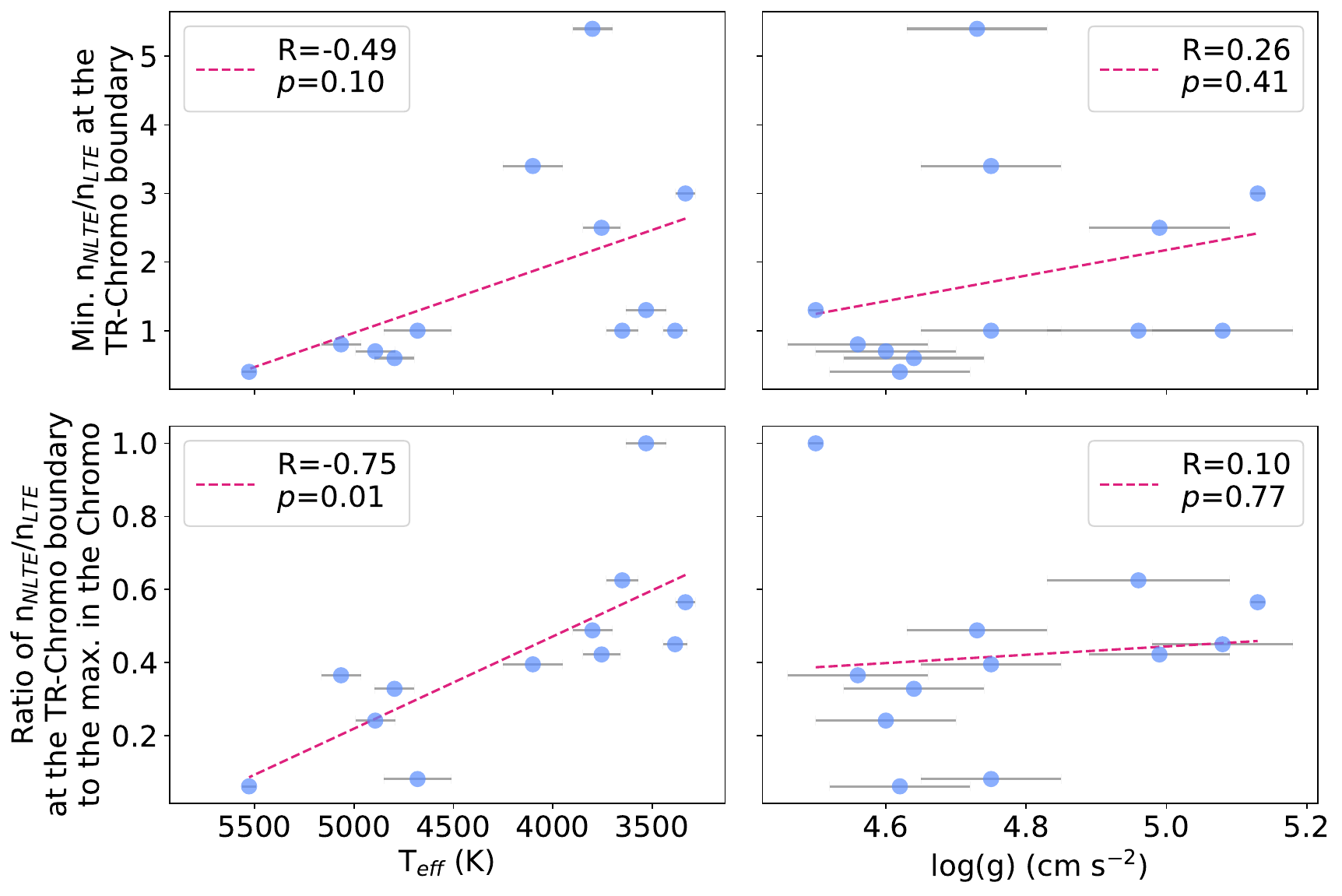}
    \caption{Minimum values for n$_{NLTE}$ /n$_{LTE}$ in the n=2 state of H I at the transition region-chromosphere boundary (\textit{left column}) and the ratio of that value to the corresponding maximum n$_{NLTE}$ /n$_{LTE}$ in the chromosphere (\textit{right column}) versus $T_{\rm eff}$ (\textit{left column}) and the logarithm of surface gravity (\textit{right column}). Linear regression fits are plotted as pink dashed lines. We find a strong trend (R = 0.75, \textit{p}=0.01) between $T_{\rm eff}$ and the ratio of the minimum n$_{NLTE}$ /n$_{LTE}$ at the transition region-chromosphere boundary to the corresponding maximum value in the chromosphere.}
    \label{fig:correlations_bi}
\end{figure}

Figure \ref{fig:correlations_bi} compares these imposed minimum values with $T_{\rm eff}$ and log($g$). We find weak correlations (R=0.49, 0.26; \textit{p}=0.1, 0.41) between the minimum n$_{NLTE}$ /n$_{LTE}$ at the transition region-chromosphere boundary and these parameters. However, when considering the ratio of the minimum value to the corresponding maximum n$_{NLTE}$ /n$_{LTE}$ in the chromosphere, the correlation with $T_{\rm eff}$ strengthens significantly (R=0.75, \textit{p}=0.01). This result provides a key step toward improving stellar modeling. The strong correlation between the ratio of minimum to maximum n$_{NLTE}$ /n$_{LTE}$ and $T_{\rm eff}$ suggests that we can apply a systematic correction to model departures from LTE. By incorporating this relationship, we can better constrain the excitation state of hydrogen in the chromosphere and transition region, leading to more accurate predictions of \lya\ emission across different stellar types.

\subsection{Identifying Trends with Flux at the Habitable Zone}
In Figure \ref{fig:correlations_p}, we find that earlier spectral types emit more \lya\ flux at the stellar surface. This suggests that planets at the same orbital distance around K stars likely experience stronger \lya-driven atmospheric photodissociation than those around M stars. We note that this trend is driven primarily by the larger radii of K stars, rather than implying that K stars are intrinsically more active than M stars. However, when scaling the \lya\ flux to the habitable zone (HZ) distance \citep{Kopparapu2014}, the \lya\ flux is comparable for both K and M stars, ranging from $\sim$2 to 20 erg cm$^{-2}$ s$^{-1}$ (Table \ref{tab:lya_flux}).  This is consistent with the findings of \citet{Richey2023}, who reported that total UV fluxes in the HZs of K and M dwarfs are broadly similar, despite differences in stellar size and surface flux.

\begin{figure}
    \centering
    \includegraphics[width=0.75\linewidth]{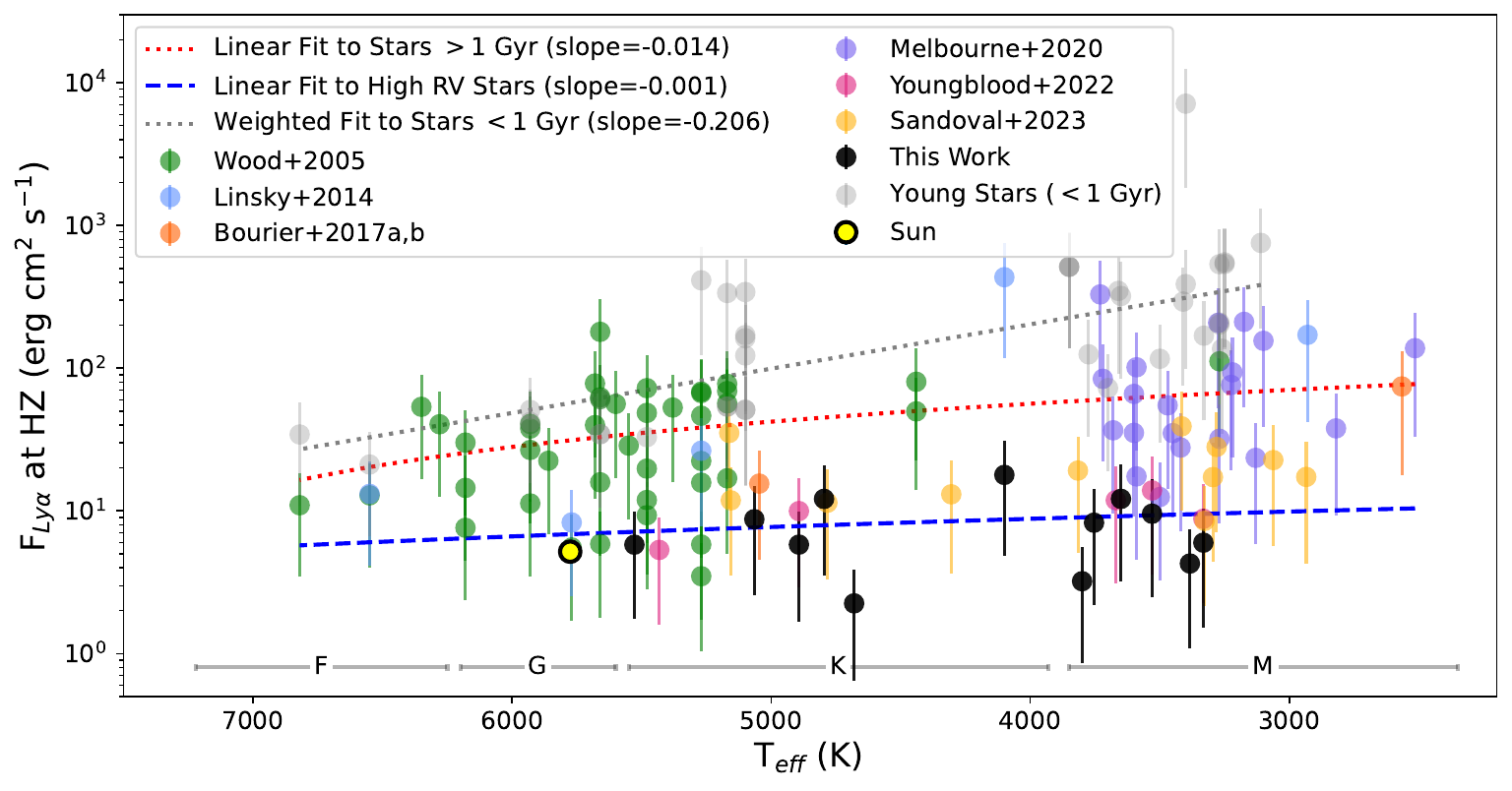}
    \caption{Reconstructed \citep{Wood2005,linsky2014,Bourrier2017a,Bourrier2017b,Melbourne2020,Sandoval2023} and directly observed \citep[This work,][]{Youngblood2022} \lya\ fluxes scaled to the habitable zone, following \cite{Kopparapu2014} and assuming a 1 Earth-mass planet. Error bars give the flux range between the Recent Venus and Early Mars limits, not the estimated uncertainties. All stars older than 1 Gyr are plotted in color, and those younger than 1 Gyr in gray. Linear regression fits to all stars $>$1 Gyr (\textit{red dotted line}), to just the high RV stars from this work and \citealt{Youngblood2022} (\textit{blue dashed line}), and a weighted fit to the young stars ($<$1 Gyr; \textit{gray dotted line}) are shown. We note that the Sun and the \citealt{Sandoval2023} values (improved \lya\ reconstructions from those presented in \citealt{Youngblood16}) align more closely with the high RV fit line.}
    \label{fig:habzone}
\end{figure}

Figure \ref{fig:habzone} illustrates the comparatively direct \lya\ measurements from this study alongside reconstructed fluxes from previous works \citep{Wood2005, linsky2014, Bourrier2017a,Bourrier2017b, Melbourne2020, Youngblood2022, Sandoval2023}, plotted against $T_{\rm eff}$ at the HZ. Since our sample is likely old (based on kinematics), we highlight stars older than 1 Gyr in color, while younger ones are shown in gray for comparison.

For the full sample of stars older than 1 Gyr, a planet in the HZ receives increasing \lya\ flux with decreasing $T_{\rm eff}$. A linear regression fit to the data suggests that a planet in the HZ of a 3000 K star receives $\approx$2.5$\times$ more \lya\ flux than a planet in the HZ of a 6000 K star. However, when isolating the high RV stars, which are likely older than 5 Gyr, the trend flattens. In contrast, among young stars ($<$1 Gyr), the \lya\ flux around later-type stars is even higher relative to earlier types, a 3000 K star emits $\approx$10$\times$ more \lya\ flux than a 6000 K star, emphasizing the role of stellar age. Notably, the high RV stars from this study, \cite{Youngblood2022}, and the Sun align along the same linear regression, suggesting that older stellar populations may follow a common evolutionary trajectory in their HZ \lya\ flux. This alignment also reinforces the reliability of the comparatively direct measurements used in this work.

Reconstructed \lya\ fluxes are widely used to correlate with spectral lines, UV photometry, and X-ray fluxes, aiding habitability assessments across stellar types. However, assumptions in \lya\ reconstructions, particularly regarding the shape of the intrinsic line profile, may introduce systematic offsets. \cite{Sandoval2023} revised earlier reconstructions from \cite{Youngblood16}, following updated methodology from \cite{Youngblood2022} with improved parameterizations of both the intrinsic line wings and the central reversal feature. Their analysis revealed that for M dwarfs, previous flux estimates were overestimated by 10--35\%, while for K stars (whose \lya\ lines exhibit deeper central reversals) the overestimates were more severe, by factors of 3--5. This disparity underscores the importance of including accurate reversal structures, especially for earlier-type stars.

Given this, it is possible that all reconstructed \lya\ fluxes may be systematically overestimated to some degree, depending on how the intrinsic line profile was treated. While many of the hotter stars in Figure~\ref{fig:habzone} already include a reversal in their reconstructions, if their fluxes are still too high, this could reflect underlying issues such as incorrect assumptions about the ISM H I column density, which strongly affects the inferred flux. However, stellar age also plays a critical role, as older stars may truly emit less \lya\ flux, so distinguishing between reconstruction error and intrinsic stellar evolution remains an important challenge.

If dwarf stars emit less intrinsic \lya\ radiation than previously thought, the radiation environments of their exoplanets may need to be revisited. However, in \cite{Peacock2022}, we explored the impact of varying \lya\ fluxes by factors of 0.35 to 1.5 on the photochemistry of high molecular weight terrestrial exoplanet atmospheres. We found that these changes produced only minor variations in spectrally active gases, and the resulting differences in atmospheric spectra would be undetectable with current observatories like JWST. This suggests that, at least for terrestrial planet atmospheres, moderate uncertainties in \lya\ flux are unlikely to significantly affect the interpretation of transmission spectra. While continued refinement of reconstruction techniques remains valuable, especially for broader stellar and planetary applications, the near-term implications for atmospheric modeling appear to be limited in scope.

\subsection{The Role of Stellar Age in Shaping \lya\ Emission}

Several studies have investigated how \lya\ emission evolves with stellar age, with implications for planetary atmospheric evolution and habitability. For instance, \citet{Johnstone2021} model the decay of X-ray, EUV, and \lya\ emission over time, finding that \lya\ decays more slowly than both X-ray and EUV luminosities. This leads to increasing $F_{\lya}/F_{\mathrm{X}}$ and $F_{\lya}/F_{\mathrm{EUV}}$ ratios with age, particularly beyond $\sim$1 Gyr. \citet{Engle2024} extend this picture observationally for M dwarfs, showing a saturation phase lasting $\sim$720 Myr for early-M dwarfs and $\sim$1.5 Gyr for mid-to-late Ms, though both phases are poorly constrained due to sparse data. Notably, \citet{Engle2024} find that \lya\ luminosity relative to bolometric luminosity is systematically higher in cooler, later-type M dwarfs, consistent with the trends reported by \citet{Linsky2020}.

Our results complement these studies by providing comparatively direct \lya\ observations for a set of 12 K and M dwarfs, with age estimates available for 10 of them. While our sample size is too small to resolve saturation-phase durations or perform subtype-separated fits, we observe general agreement with the evolutionary patterns described above. Among stars older than 1 Gyr, the \lya\ flux at the HZ decreases with increasing $T_{\rm eff}$, suggesting that lower-mass stars remain relatively \lya-active for longer. Moreover, our subset of high radial velocity stars (likely all older than 5 Gyr) lies along a tight \lya–$T_{\rm eff}$ relation, consistent with an evolved population that may follow a common decay track. These findings echo the late-time convergence seen in \citet{Johnstone2021} and reinforce the interpretation that \lya\ emission persists as a significant UV source even in old, low-mass stars.

{To further investigate age evolution, we compared our observationally constrained $L_{\mathrm{Ly}\alpha}/L_{\mathrm{bol}}$ ratios to the empirical relations from \citet{Engle2024} for M dwarfs (Figure \ref{fig:age}). Our three M0--M2 stars fall within or exceeding the upper edge of the model's $1\sigma$ envelope at ages $>5$ Gyr, with a possible trend of increasing $L_{\mathrm{Ly}\alpha}/L_{\mathrm{bol}}$ with age across these objects. In contrast, our two mid-M stars (L 802-6 and Barnard's Star) lie significantly below the empirical M2.5--M6.5 relation, even after accounting for model uncertainties. This trend is in tension with the conclusions of \citet{Engle2024} and \citet{Linsky2020}, who find that $L_{\mathrm{Ly}\alpha}/L_{\mathrm{bol}}$ increases toward cooler, later-type M dwarfs. Our data suggest instead that, at least among older field stars, early-M dwarfs may retain stronger relative \lya\ emission than their mid-M counterparts. This discrepancy could reflect intrinsic variability, sampling effects, or limitations in the \lya\ reconstructions at late ages.

\begin{figure}
    \centering
    \includegraphics[width=0.5\linewidth]{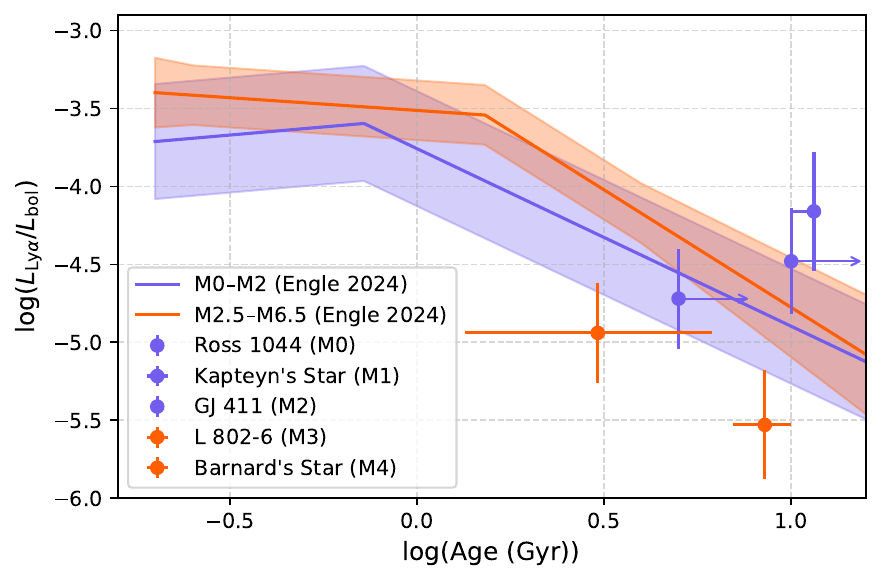}
    \caption{The evolution of $\log(L_{\mathrm{Ly}\alpha}/L_{\mathrm{bol}})$ as a function of stellar age for M dwarf stars. Solid lines show the parameterized empirical trends for M0--M2 and M2.5--M6.5 spectral types from \citet{Engle2024}, with shaded regions representing $1\sigma$ uncertainties on the fits. Overplotted are our sample stars, color-coded by spectral type: M0--M2 (purple) and M3--M4 (orange). The three M0--M2 stars lie within or exceeding the upper bound of the extrapolated uncertainty envelope at older ages. In contrast, the M3--M4 stars fall significantly below the empirical track, suggesting either lower intrinsic activity or potential limitations of the model extrapolation at late ages for mid-M dwarfs.
}
    \label{fig:age}
\end{figure}

\section{Conclusions}
We present \lya\ profiles for 12 high radial velocity stars, with $\approx$50–95\% of their intrinsic profiles directly observable. This sample uniformly covers the critical K to M $T_{\rm eff}$ range, allowing us to identify the point where self-reversals in \lya\ profiles become significant ($T_{\rm eff} >$ 4000 K) and to reduce systematic uncertainties in future profile reconstructions. We find that \lya\ reversal depth correlates with $T_{\rm eff}$—as $T_{\rm eff}$ decreases, self-reversals become less pronounced, with M stars exhibiting little to no reversal.

Two stars, Ross 1044 and Ross 451, were observed over multiple days, revealing $\sim$20\% variability. For Ross 1044, the self-reversal depth fluctuated by 5\% over a timescale of less than a week, whereas Ross 451, which lacks a reversal, showed no change in peak-to-trough ratio. In both cases, variability was localized to the line core, mirroring solar behavior. This suggests that \lya\ variability in other systems may be underestimated when the core is fully absorbed—if only the wings are observable, they may appear stable while significant variations occur in the unseen core. Given that these two stars are old, this effect may be even more pronounced in younger, more active stars, similar to the Sun, which exhibits up to 155\% variation over its 11-year cycle.

Beyond variability, we examined the relationship between \lya\ flux, atmospheric structure, and stellar parameters. We find strong correlations between \lya\ flux, peak-to-trough ratio, and the departure coefficient, n$_{NLTE}$ /n$_{LTE}$, in the n=2 state of H I with $T_{\rm eff}$. These trends provide critical constraints for refining stellar atmosphere models, particularly for K and M dwarfs, where direct measurements of intrinsic \lya\ profiles remain difficult.

Comparing reconstructed \lya\ flux in the HZ of typical stars and high radial velocity stars reveals a systematic offset: among stars older than 1 Gyr, the high-RV stars show consistently higher flux, with a mild trend indicating that M stars may have stronger HZ \lya\ flux than G and K stars. However, for high radial velocity stars, the trend is flat. This offset may stem from differences in age, as high radial velocity stars are likely older ($>$5 Gyr), or from overestimated reconstructed fluxes. The latter is supported by the Sun's directly measured value and improved reconstructions of M and K stars from \citealt{Sandoval2023}, which align more closely with the high radial velocity sample.

Finally, we compared $L_{\mathrm{Ly}\alpha}/L_{\mathrm{bol}}$ for the M dwarfs in our sample to existing empirical estimates from \citet{Engle2024}. Our results suggest that early-M stars maintain higher relative \lya\ emission at late ages than mid-M stars, contradicting the trend of increasing activity toward later spectral types reported in previous studies. Due to the scarcity of high radial velocity stars enabling intrinsic \lya\ profile reconstruction, this dataset likely represents the most complete observational sample possible, emphasizing the importance of developing refined models to fully capture M dwarf chromospheric evolution.

Overall, this work provides key observational benchmarks for \lya\ emission across K and M stars, reducing uncertainties in reconstructions and informing models of stellar atmospheres and exoplanet environments. Future observations of these stars, particularly to better quantify short- and long-term variability, will further refine our understanding of stellar UV radiation and its impact on exoplanet atmospheres and habitability.

\begin{acknowledgments}

We thank the anonymous referee for their thoughtful comments and suggestions, which improved the quality of this manuscript. This research was supported by HST-GO-16646. S.P. and E.L.S. acknowledge support from the CHAMPs (Consortium on Habitability and Atmospheres of M-dwarf Planets) team, supported by the National Aeronautics and Space Administration (NASA) under grant nos. 80NSSC21K0905 and 80NSSC23K1399 issued through the Interdisciplinary Consortia for Astrobiology Research (ICAR) program.
S.P. also acknowledges support from NASA under award number 80GSFC24M0006.

\end{acknowledgments}

\appendix

\cite{Peacock2022} found that the depth of the self-reversal in the \lya\ line directly depends on the ratio of the non-local thermodynamic equilibrium (NLTE) to LTE number density (n$_{NLTE}$ /n$_{LTE}$) for hydrogen in the n=2 state near the transition region-chromosphere boundary. Our initial model successfully reproduced the observed \lya\ line widths but severely underestimated the line core due to a sharp decrease in this region. To match the full line profile, we imposed minima on the n=2 level population of H I at the chromosphere-transition region boundary. The following figures compare the initial and improved models, showing n$_{NLTE}$ /n$_{LTE}$ versus column mass and the computed \lya\ profiles against observations.

\begin{figure}
    \centering
    \includegraphics[width=0.75\linewidth]{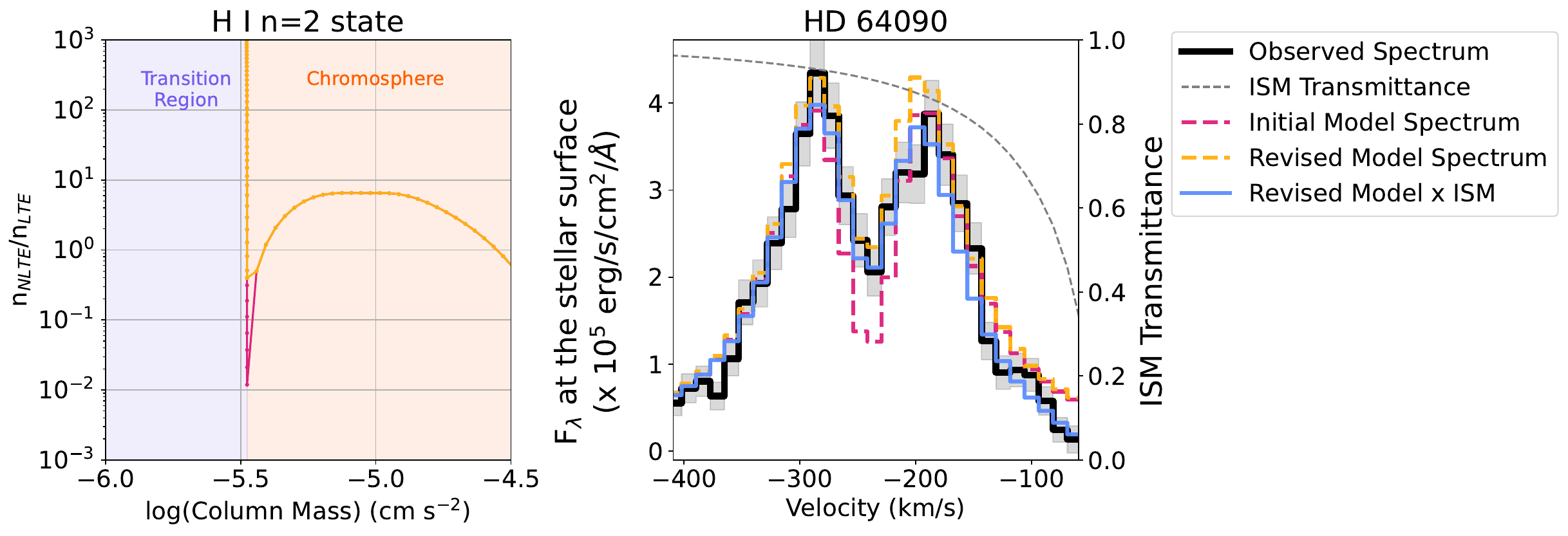}
    \includegraphics[width=0.75\linewidth]{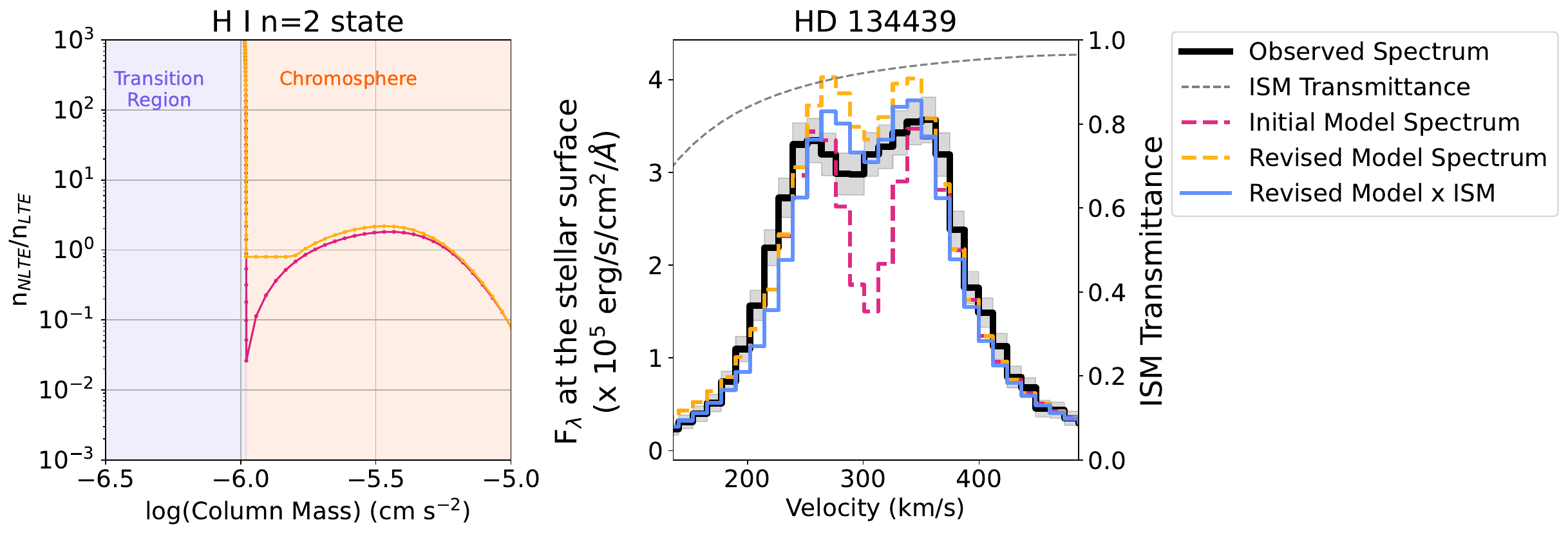}
    \includegraphics[width=0.75\linewidth]{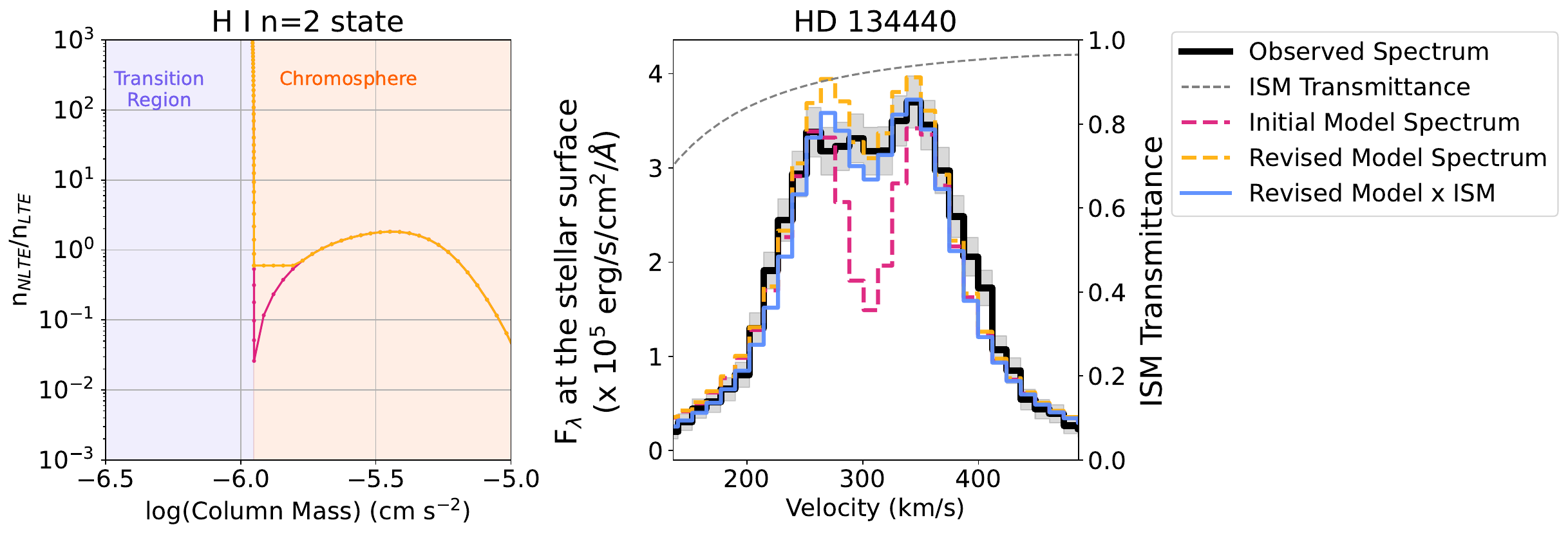}
    \includegraphics[width=0.75\linewidth]{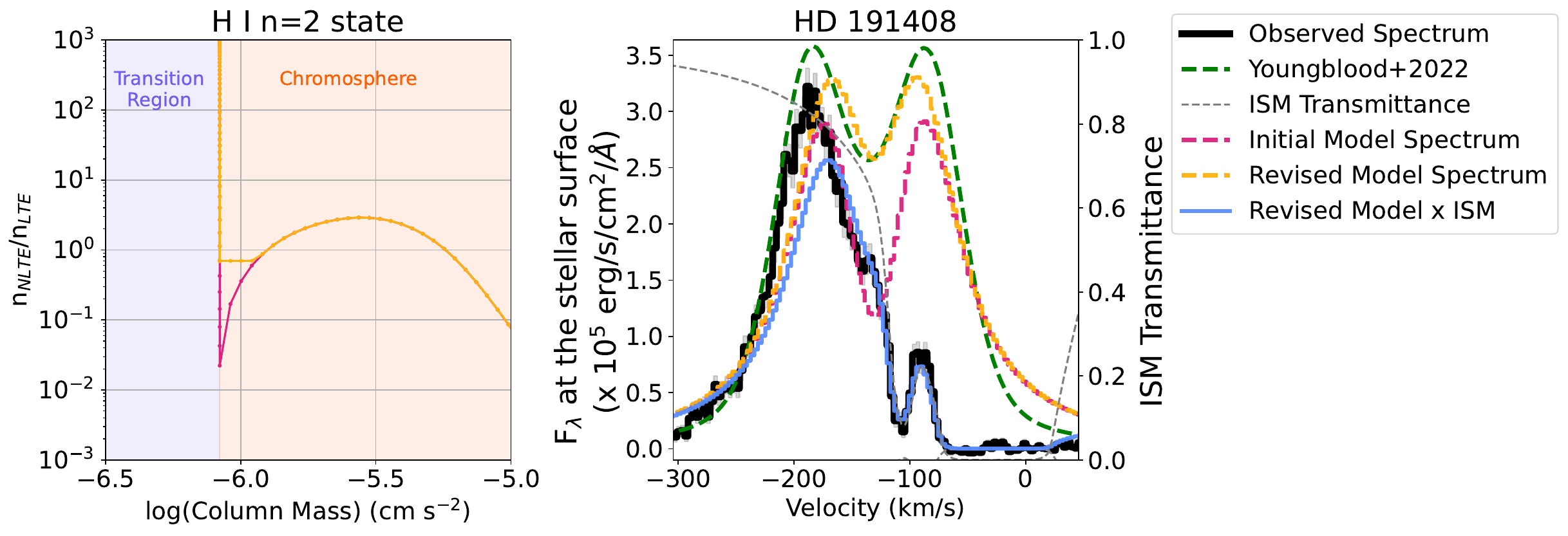}
    \caption{\textit{Left}: n$_{NLTE}$ /n$_{LTE}$ for hydrogen in the n=2 state versus column mass zoomed in to show detail in the transition region and upper chromosphere (these values go to 1 deep in the atmosphere). Initial models are shown in pink, revised models with an imposed minimum set at the boundary between these layers is shown in orange. \textit{Right}: \lya\ profiles. HST/STIS observations are plotted in black, with associated errors shaded in gray. The ISM transmittance curve for each star is plotted as a gray dashed line. Initial intrinsic \texttt{PHOENIX} model profiles are plotted in pink and show deep central reversals. Revised intrinsic \texttt{PHOENIX} model profiles are plotted in orange. The revised intrinsic \texttt{PHOENIX} model, after being multiplied by the ISM transmittance curve, is plotted in blue — this profile should match the observations. Stars shown in this figure are (from top to bottom): HD 64090 (K0), HD 134439 (K1), HD 134440 (K2), and HD 191408 (K2.5). We include a comparison to the intrinsic \lya\ profile computed in \citealt{Youngblood2022} for HD 19148, plotted in green.}
    \label{fig:a1}
\end{figure}

\begin{figure}
    \centering
    \includegraphics[width=0.75\linewidth]{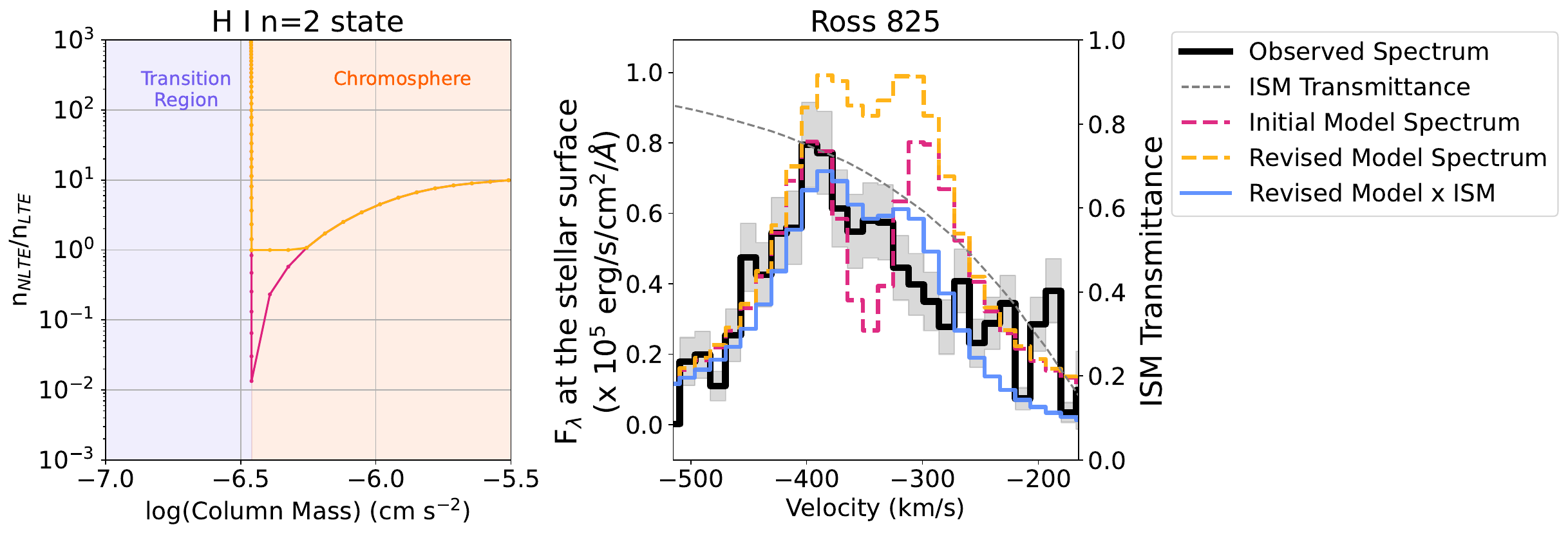}
    \includegraphics[width=0.75\linewidth]{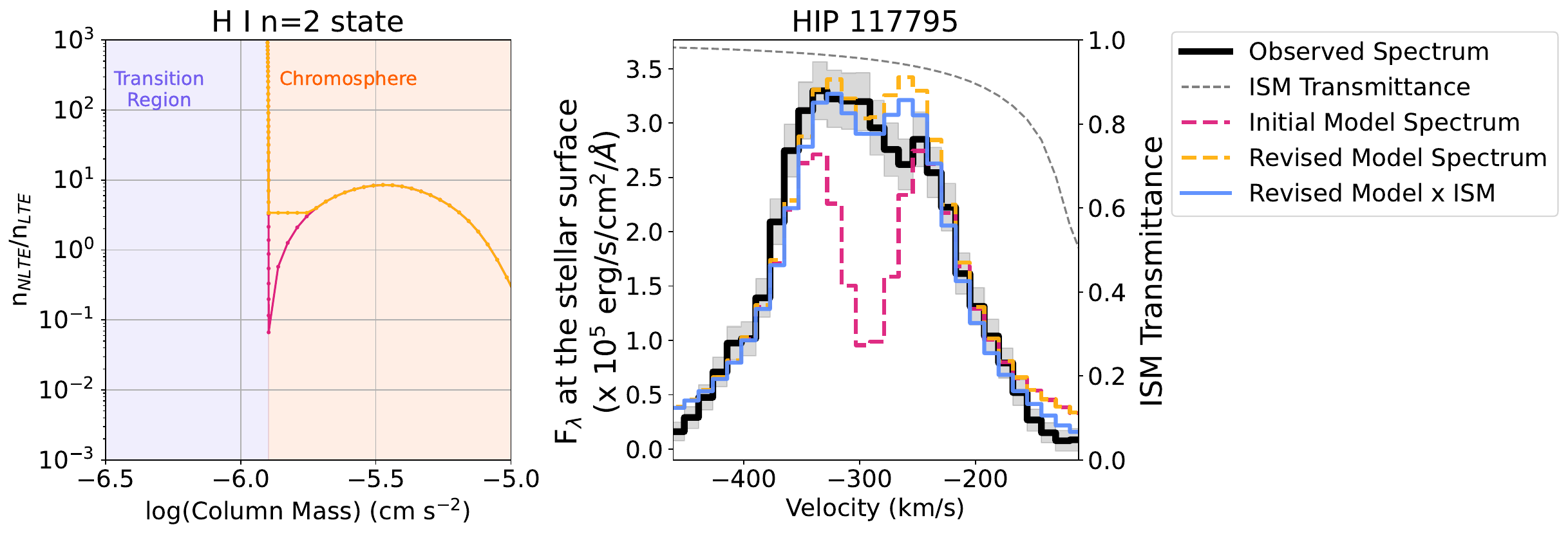}
    \includegraphics[width=0.75\linewidth]{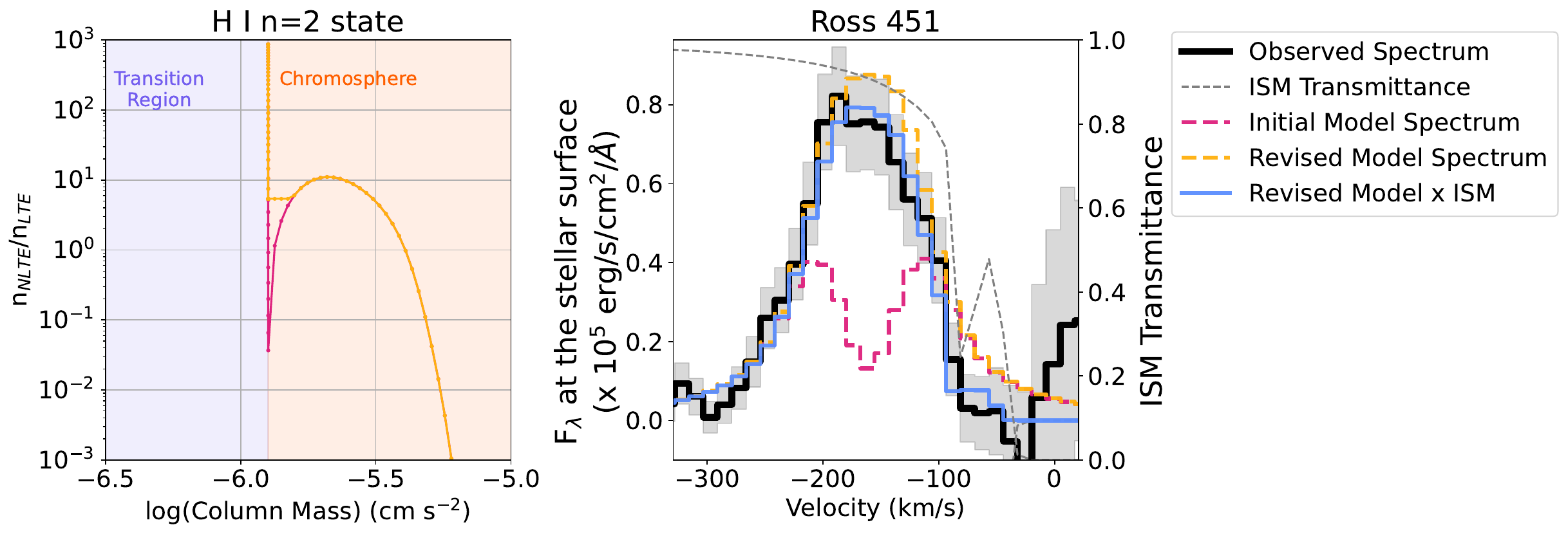}
    \includegraphics[width=0.75\linewidth]{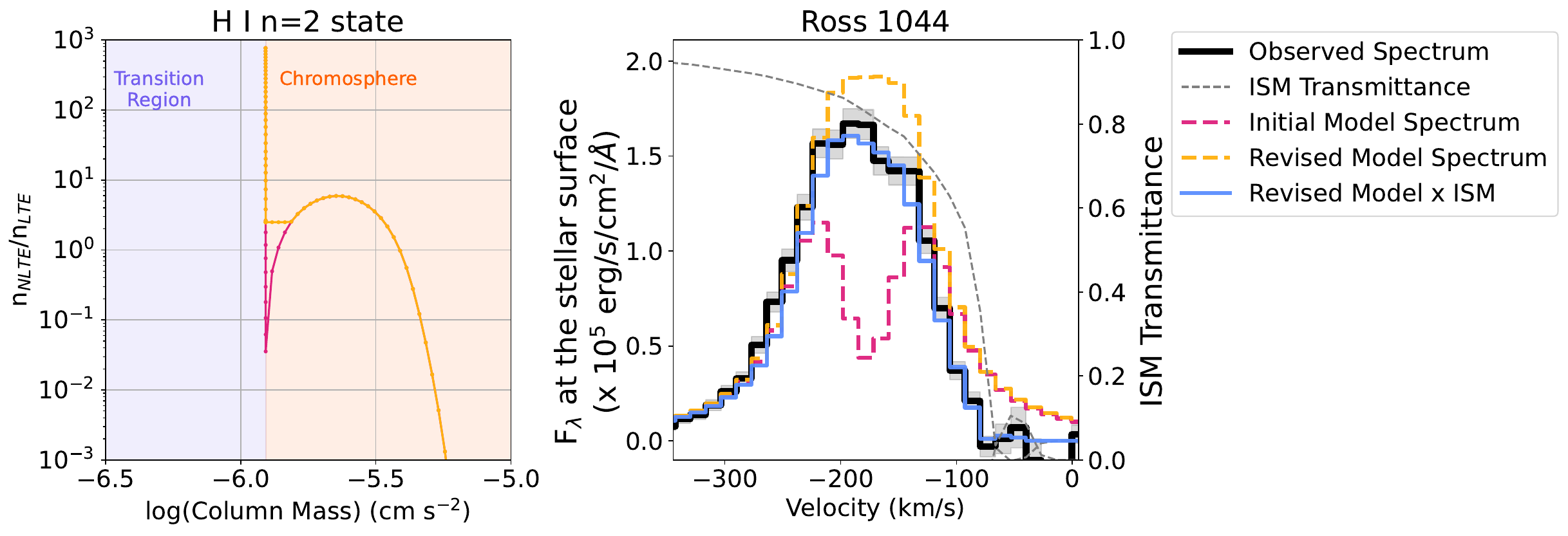}
    \caption{\textit{Same as Figure \ref{fig:a1}}. Stars shown in this figure are (from top to bottom): Ross 825 (K3), HIP 117795 (K8), Ross 451 (M0), and Ross 1044 (M0).}
    \label{fig:a2}
\end{figure}

\begin{figure}
    \centering
     \includegraphics[width=0.75\linewidth]{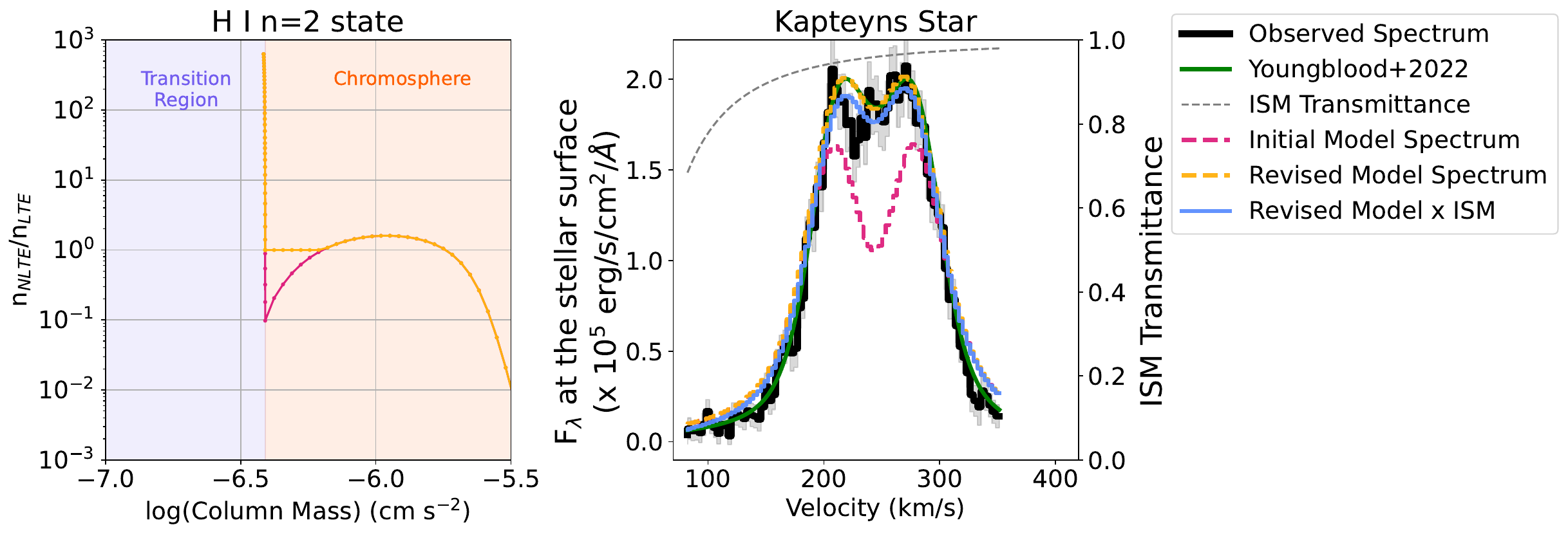}
    \includegraphics[width=0.75\linewidth]{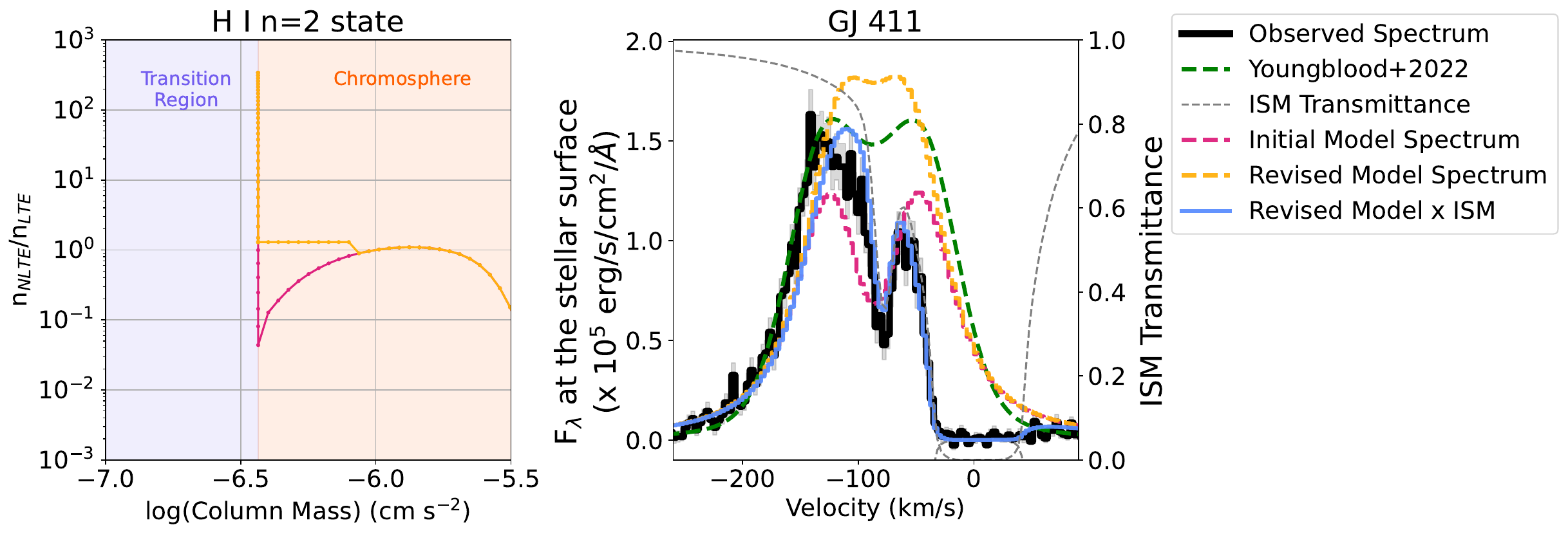}
    \includegraphics[width=0.75\linewidth]{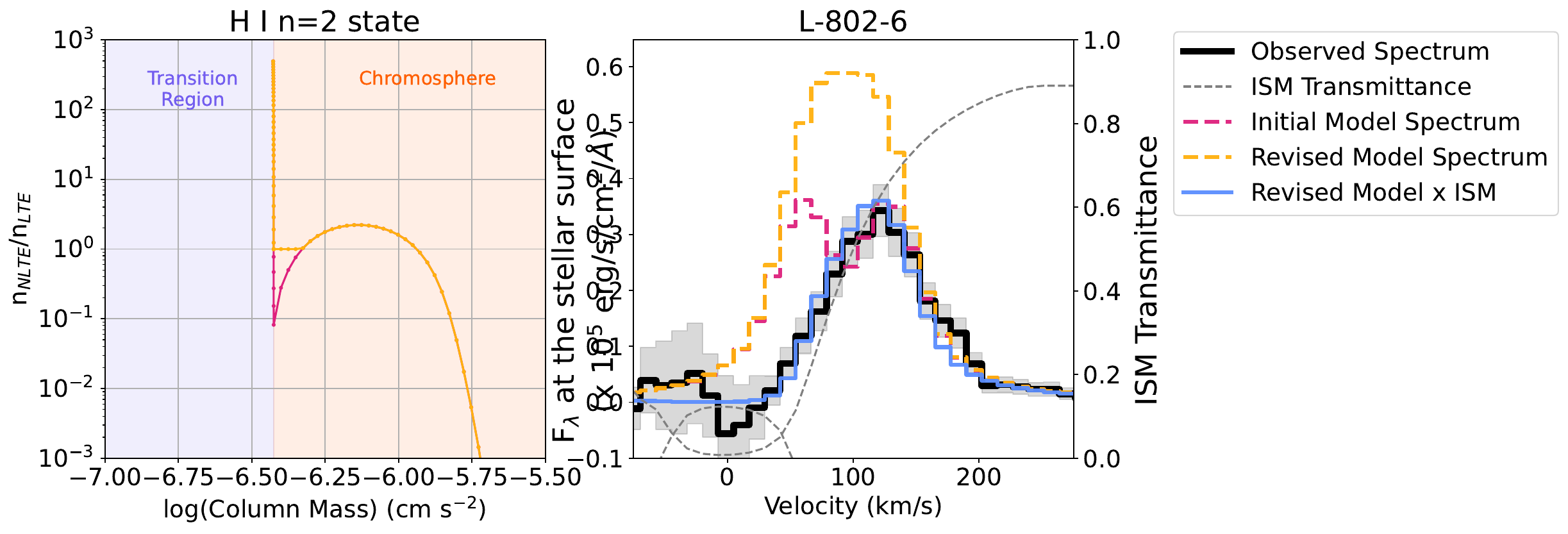}    
    \includegraphics[width=0.75\linewidth]{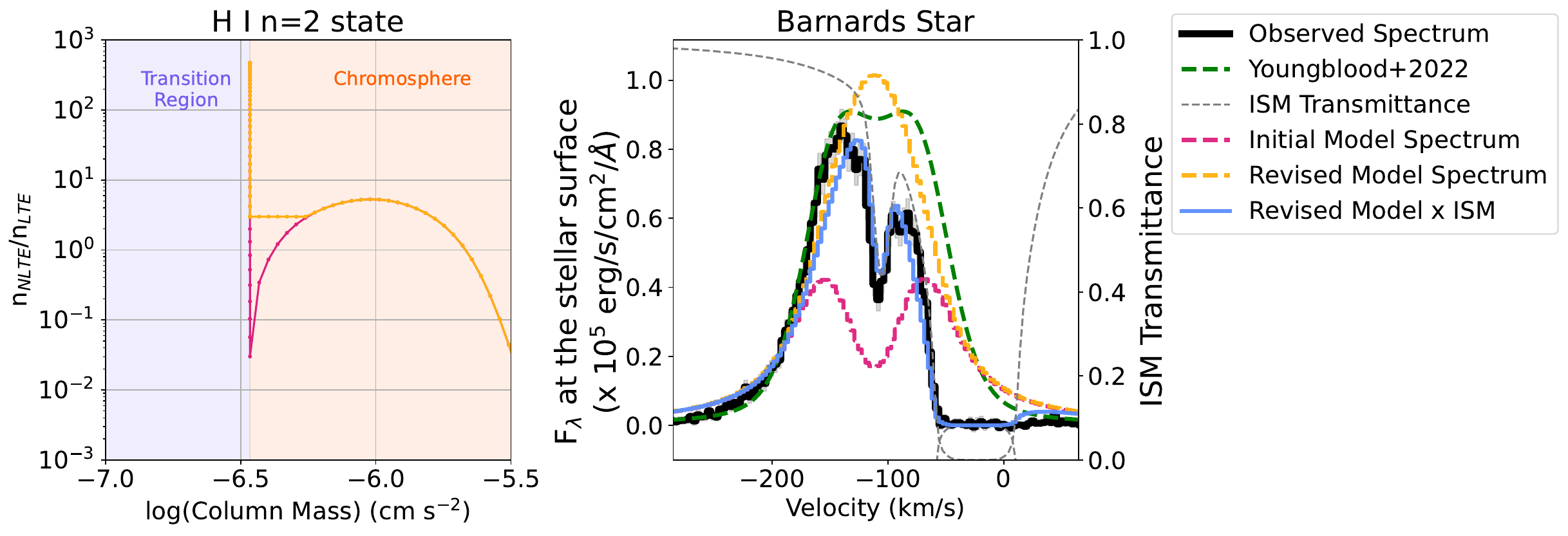}
    \caption{\textit{Same as Figure \ref{fig:a1}.} Stars shown in this figure are (from top to bottom): Kapteyn's Star (M1), GJ 411 (M2), L-802-6 (M3), Barnard's Star (M4). We include comparisons to the intrinsic \lya\ profiles computed in \citealt{Youngblood2022} for Kapteyn's Star, GJ 411, and Barnard's Star, plotted in green.}
    \label{fig:a3}
\end{figure}

\vspace{5mm}
\facilities{HST(STIS)}

\end{document}